\documentclass[12pt]{article}
\usepackage[utf8]{inputenc} 
\usepackage[english]{babel} 
\usepackage{lipsum} 
\usepackage{url} 
\usepackage[dvipsnames]{xcolor}
\usepackage{authblk}
\usepackage{tcolorbox}
\usepackage{float}
\usepackage{mathtools}
\usepackage{booktabs} 
\usepackage{fancyhdr} 
\usepackage{hyperref} 
\usepackage{amsmath} 
\usepackage[toc,page]{appendix}
\usepackage{bbold}
\usepackage{blindtext}
\usepackage{caption}
\usepackage{subcaption}
\usepackage{amsthm} 
\usepackage{mathtools} 
\usepackage{amsthm} 
\usepackage{bm} 
\usepackage{braket} 
\usepackage{tikz} 
\usepackage{amssymb} 
\usepackage{amsfonts} 
\newcommand{\R}{\mathbb{R}}

\newcommand{\de}{\partial}
\newcommand{\diff}{\mathrm{d}}

\newcommand{\be}{
\begin{equation}
}

\newcommand{\beLabel}[1]{
\begin{equation}
\label{eq:#1}
}

\newcommand{\ee}{
\end{equation}
}

\newcommand{\eeLabel}{
\end{equation}
}

\let\oldref\ref
\renewcommand{\ref}[1]{(\oldref{#1})}

\title{\textbf{
\vskip-3cm
On-Shell Flow}}
\date{}
\author[a]{Davide De Biasio}
\affil[a]{Max--Planck--Institut f\"ur Physik, Werner--Heisenberg--Institut, \newline
F\"ohringer Ring 6, 80805 M\"unchen, Germany}

\usepackage{cite}
\numberwithin{equation}{section}
\begin{document}
\fancypagestyle{plain}{%
	\fancyhead[R]{MPP-2022-134}
	\renewcommand{\headrulewidth}{0pt}
}
\maketitle
\begin{center}Abstract:
\end{center}
In this work, the problem of constructing geometric flow equations that preserve Einstein field equations for the spacetime metric is addressed. After having briefly discussed the main features of Ricci flow, the on-shell flow equations for a system comprised of a dynamical metric and a set of matter fields are constructed. Then, two examples in which the matter content is just a single, self-interacting scalar field are analysed in detail, imposing the geometric flow equations to be the action-induced ones. In conclusion, an explicit connection to the Swampland Distance Conjecture is proposed.
\newpage
\tableofcontents
\newpage
\section{Introduction}
In the last decade, the Swampland Program \cite{vafa2005string,Palti:2019pca,vanBeest:2021lhn,Brennan:2017rbf} has vigorously substantiated the long-standing intuition that any viable attempt to merge general relativistic gravity and particle physics, even in a low-energy regime in which explicit quantum gravity effects can be neglected, would have pushed our inquiry beyond the domain in which standard quantum field theory techniques can be trusted and reliably applied. It could indeed be argued that there is now overwhelming evidence suggesting that the ordinary effective field theory approach, in which one is simply allowed to integrate out degrees of freedom lying above an energy cut-off $\Lambda_{\text{eff}}$ and disregard high-energy effects, fails when quantum fields are coupled to a dynamical space-time metric. The essential observation around which the Swampland Program developed and flourished is therefore that gravity non-trivially constraints the features of low-energy effective theories, in ways which are not captured by the traditional perspective. Such restrictions are often stated in the form of the so-called Swampland Conjectures. It is thus reasonable to consider the set of apparently consistent quantum field theories coupled to general relativity below a given energy scale and partition it in two subsets:
\begin{itemize}
    \item The Landscape, comprised of those theories which are consistent with the restrictions imposed by gravity and hence admit an ultraviolet completion to quantum gravity.
    \item The Swampland, comprised of those theories which, albeit looking well-behaved at first glance, are not consistent with the restrictions imposed by gravity and hence do not admit an ultraviolet completion to quantum gravity.
\end{itemize}
Further exploring this line of thought, it seems reasonable to model the above-mentioned set as some kind of moduli space. Namely, we regard it as a (perhaps extremely complicated) geometric object, charted by a set of generalised moduli. One of the first and best-established statements concerning the properties of such moduli space is the Swampland Distance Conjecture \cite{Ooguri:2006in}, in which it is argued that large displacements in the moduli should correspond to the appearance of an infinite tower of light states in the low energy spectrum of the theory. 
Regarding the space-time metric as a generalised modulus of the theory and attempting to find a natural way of producing displacements in its components, inspiration was drawn from the graviton renormalization group flow in string theory $\sigma$-models. By doing so, the Distance Conjecture was then extended and specified in terms of various geometric flow equations \cite{Kehagias:2019akr,Bykov:2020llx,Luben:2020wix,DeBiasio:2020xkv,de2022geometric,DeBiasio:2022omq}, in which an auxiliary flow parameter $\lambda$ was introduced and a $\lambda$-evolution of the metric got sourced by an expression involving not only the metric itself and its derivatives, often combined in order to form the associated Ricci tensor, but also any other field appearing in the theory, along with its covariant derivatives. Generalising the work of A.M.Polyakov on the renormalisation of the $O\left(N\right)$-invariant nonlinear $\sigma$-model in $2+\epsilon$ dimensions \cite{Polyakov:1975rr}, geometric flow techniques were first introduced in the context of string theory and condensed matter physics by D.Friedan \cite{friedan1980nonlinear,friedan1985nonlinear}. If interested in some technical monographs covering the methods applied in our derivations, or at least in the corresponding Euclidean discussions, the reader is strongly suggested to refer to \cite{chow2004ricci,Cao2006ACP,morgan2007ricci,kleiner2008notes}, together with Hamilton's foundational work \cite{hamilton1982} and the remarkable application of Ricci flow in Perelman's proof of the geometrization conjecture \cite{Perelman:2006un}. In that context, the paradigmatic flow equations for a metric and a scalar appeared as:
\begin{equation}
        \begin{cases}
    \diff_{\lambda}g_{\mu\nu}=-2R_{\mu\nu}-2\nabla_\mu\nabla_\nu\phi\ ,\\
    \diff_{\lambda}\phi=-R-\Delta \phi\ ,\\
    g_{\mu\nu}\left(0\right)=\Bar{g}_{\mu\nu}\ ,\\
    \phi\left(0\right)=\Bar{\phi}\ ,
    \end{cases}
\end{equation}
Given a theory of fields coupled to gravity and a set of flow equations for both the metric and the fields, together with an on-shell initial configuration for the flow, there is no guarantee that the $\lambda$-evolution will produce on-shell configurations, even for very small values of the flow parameter. In general, instead, we observe a deviation from the equations of motion of the theory as soon as we switch-on the flow. In the following work, we will try to construct on-shell flows, preserving at least part of the equations of motion of the theories under consideration. By doing so, we will be able to regard the induced $\lambda$-evolution as a path in the generalised moduli space of actual solutions of the equations of motion. If this will only be reasonably achieved for a subset of the fields, it will nevertheless be fruitful to project the path produced by the flow to the moduli space directions charted by such fields and obtain some kind of partially on-shell flow.

\newpage
\section{Definition of the theory} \label{theory}
In this section, the general form of the theories considered in our work will be outlined in detail.
Let $\mathcal{M}$ be a $D$-dimensional smooth manifold, endowed with a pseudo-Riemannian metric tensor $g_{\mu\nu}$, and $\varphi_A:\mathcal{M}\rightarrow\mathcal{C}_A$ be a generic family of $N$ matter fields on $\mathcal{M}$, together with the corresponding manifolds $\mathcal{C}_A$ in which they take values. If $N\geq 1$, we clearly have $A=1,\dots, N$.
Furthermore, let $\mathcal{L}_M\left[g_{\mu\nu},\varphi_A\right]$ be the Lagrangian describing matter dynamics on $\mathcal{M}$. Therefore, by imposing the space-time metric behaviour to be fixed by $D$-dimensional general relativity with a cosmological constant $\Lambda$, the full action becomes:
\begin{equation}\label{action}
    \mathcal{S}\equiv\frac{1}{2\kappa}\int_{\mathcal{M}}\diff^D x\sqrt{-g}\left(R-2\Lambda\right)+\int_{\mathcal{M}}\diff^D x\sqrt{-g}\mathcal{L}_M\ 
\end{equation}
If we impose $N=0$ and $\Lambda=0$, the above theory simply reduces to standard $D$-dimensional vacuum general relativity. Considering, instead, the most general case captured by \eqref{action}, the equations of motion for the space-time metric $g_{\mu\nu}$ can be explicitly written down as:
\begin{equation}\label{eom}
        G_{\mu\nu}+g_{\mu\nu}\Lambda=\kappa T_{\mu\nu}\ ,
    \end{equation}
    where the Einstein tensor $G_{\mu\nu}$ is defined as
    \begin{equation}
        G_{\mu\nu}\equiv R_{\mu\nu}-\frac{1}{2}g_{\mu\nu}R\ ,
    \end{equation}
    and the energy-momentum tensor has the form:
    \begin{equation}
        T_{\mu\nu}=\frac{-2}{\sqrt{-g}}\frac{\delta\left(\sqrt{-g}\mathcal{L}_M\right)}{\delta g^{\mu\nu}}\ .
    \end{equation}
    
    The strength of the coupling among the matter fields and space-time curvature is controlled by the constant $\kappa$, which plays the role of a rescaled $D$-dimensional Newton constant. As was briefly outlined at the end of \cite{DeBiasio:2022omq}, one might consider the case in which the coupling constants themselves are subject to some kind of non-trivial flow. This would perhaps allow for a connection, or at least a comparison, between our kind of geometric flows and the more understood renormalization group flow for specific theories of the form \eqref{action}. In this regard, the most interesting and significant case would be the one in which the full flow behaviour could somehow be moved to the coupling constants. Even if such perspectives might be extremely interesting to explore, they go way beyond the scope of this work and will not be considered in the next sections. They would nevertheless make a natural candidate for a future development of our analysis.
\section{Remarks on Ricci flow}\label{rfsec}
The scope of the following section is by no means that of developing a complete discussion of the features and properties of Ricci flow, let alone doing so with the level of rigour usually expected from a technical review on the subject matter. Once more, the mathematically oriented reader is encouraged to consult the references and monographs \cite{chow2004ricci,Cao2006ACP,morgan2007ricci,kleiner2008notes,hamilton1982,Perelman:2006un}. Here we will rather focus on those aspects of Ricci flow theory which will have direct and relevant implications for the rest of our work. Specifically, we will briefly comment on the well-posedness of some Ricci flow generalisations and on the particular care that must be taken when applying such evolution equations to Lorentzian metrics, while most of the usual derivations examine Euclidean manifolds. 
\subsection{Definition of the flow}
As the one introduced at the beginning of section \ref{theory}, let $\mathcal{M}$ be a $D$-dimensional smooth manifold which can be endowed with a pseudo-Riemannian metric with Lorentzian signature $\left(-,+,\dots,+\right)$. Inspired by the formal discussion developed in \cite{1992math......1259G} for the analogous Euclidean case, we consider the, perhaps infinite-dimensional, manifold $\mathcal{G}_\mathcal{M}$ of all possible pseudo-Riemannian metrics on $\mathcal{M}$. From this perspective, any point of $\mathcal{G}_\mathcal{M}$ corresponds to a different expression for the $D$-dimensional metric $g_{\mu\nu}$ of $\mathcal{M}$. In general, one could endow $\mathcal{G}_\mathcal{M}$ itself with an appropriate metric and use it to compute geodesic distances among different metrics on $\mathcal{M}$. In the context of string theory, this was widely explored in \cite{Kehagias:2019akr} and \cite{Bonnefoy:2019nzv}. As far as the following work is concerned, it is enough to stress the fact that any continuous $1$-parameter family of metrics $g_{\mu\nu}\left(\lambda\right)$ on $\mathcal{M}$ can be regarded as a continuous path in $\mathcal{G}_{\mathcal{M}}$. In particular, this applies to the case of Ricci flow and generalisations thereof. Considering the former, which is also the most studied and best-understood, we have that a family of metrics $g_{\mu\nu}\left(\lambda\right)$ on a $D$-dimensional smooth manifold $\mathcal{M}$, where $\lambda$ is a positive real flow parameter, is said to evolve according to Ricci flow from an initial condition $\Bar{g}_{\mu\nu}$ if
\begin{equation}\label{ricciflow}
\frac{\de g_{\mu\nu}\left(\lambda\right)}{\de\lambda}=-2R_{\mu\nu}\left(\lambda\right)
\end{equation} 
for any $\lambda\geq 0$, where $R_{\mu\nu}\left(\lambda\right)$ is the Ricci tensor associated to $g_{\mu\nu}\left(\lambda\right)$ and:
\begin{equation}
g_{\mu\nu}\left(0\right)=\Bar{g}_{\mu\nu}\ .
\end{equation}
It must be stressed that we can either have $\lambda\in\left[0,\infty\right)$ or $\lambda\in\left[0,\lambda_{\text{max}}\right)$, with $\lambda_{\text{max}}\in\R$, depending on whether the flow encounters a singularity or not. This aspect is extremely important when trying to study flow asymptotics, which might require surgery procedures to be applied in case a singularity in indeed encountered, but will not be relevant for the current discussion. 
Since the theory must be invariant under $\lambda$-dependent diffeomorphisms, we can introduce a generalisation
\begin{equation}\label{riccidetflow}
\frac{\de g_{\mu\nu}\left(\lambda\right)}{\de\lambda}=-2R_{\mu\nu}\left(\lambda\right)+\nabla_{(\mu}\xi_{\nu)}\left(\lambda\right)\ ,
\end{equation} 
of \eqref{ricciflow}, where $\xi_{\nu}\left(\lambda\right)$ is a $\lambda$-dependent vector field generating a diffeomorphism. Equation \eqref{riccidetflow} is typically called Ricci-DeTurk flow. As was phrased in slightly different terms in \cite{Headrick:2006ti}, \eqref{ricciflow} and \eqref{riccidetflow} produce different paths in $\mathcal{G}_{\mathcal{M}}$, but the same path in the quotient of $\mathcal{G}_{\mathcal{M}}$ over point-wise defined diffeomorphisms. Considering that diffeomorphisms are gauge in general relativity, we expect the latter to be more appropriate for discussing the properties of our low energy effective theories. By introducing three real parameters $\alpha$, $\beta$ and $\gamma$, the expression \eqref{riccidetflow} can be further generalised to
\begin{equation}\label{genricciflow}
\frac{\de g_{\mu\nu}}{\de\lambda}=-\alpha R_{\mu\nu}+\beta R g_{\mu\nu}+\gamma g_{\mu\nu}+\nabla_{(\mu}\xi_{\nu)}\ ,
\end{equation} 
where the $\lambda$-dependence was dropped (and will be dropped from now on) for the sake of simplicity. When $\gamma$ is set equal to zero, the resulting flow is usually referred to as Ricci-Bourguignon flow: 
\begin{equation}\label{riccibouflow}
\frac{\de g_{\mu\nu}}{\de\lambda}=-\alpha R_{\mu\nu}+\beta R g_{\mu\nu}+\nabla_{(\mu}\xi_{\nu)}\ .
\end{equation} 
A broad discussion of its properties can be found in \cite{catino2017ricci}. This is the assumption we will too make from now on. Setting aside for a moment the problems that working with a Lorentzian signature might produce, we still do not know whether an equation of the form \eqref{riccibouflow} can define a well-posed set of differential equations. We will precisely address such issue in subsection \ref{wellposed}, by restricting the possible values of $\alpha$ and $\beta$, so that pathologies and instabilities are excluded.
\subsection{Well-posedness of the flow}\label{wellposed}
In order to investigate whether a differential equation of the form \eqref{genricciflow} is well-posed or not, we will follow the discussion developed in \cite{Headrick:2006ti}. For the moment being, we perform the typical analysis one would carry on in Euclidean signature and postpone, as we anticipated previously, any comment on the difficulties that working with Lorentzian signature might produce. Therefore, we linearise $g_{\mu\nu}$ around flat space-time as
\begin{equation} 
g_{\mu\nu}\left(\lambda\right)\approx\eta_{\mu\nu}+\epsilon h_{\mu\nu}\left(\lambda\right)\ ,
\end{equation} 
with $0<\epsilon\ll 1$, and consider the leading order expansion in $\epsilon$. This way, we can define $h\equiv\eta^{\mu\nu}h_{\mu\nu}$ and get:
\begin{equation}\label{fort}
R_{\mu\nu}\approx  \frac{\epsilon}{2}(\partial_\sigma\partial_\mu h^\sigma_\nu + \partial_\sigma\partial_\nu h^\sigma_\mu - \partial_\mu\partial_\nu h - \Delta h_{\mu\nu})\ .
\end{equation}
Concerning the Ricci scalar, we instead get:
\begin{equation}\label{fors}
R\approx  \epsilon\left(\partial_\mu\partial_\nu h^{\mu\nu} - \Delta h\right)\ .
\end{equation} 
By neglecting the diffeomorphism and considering the above expansions, Ricci-Bourguignon flow \eqref{riccibouflow} approximates to:
\begin{equation}
\begin{split}
\frac{\de h_{\mu\nu}}{\de\lambda}=\ &-\frac{\alpha}{2}(\partial_\sigma\partial_\mu h^\sigma_\nu + \partial_\sigma\partial_\nu h^\sigma_\mu - \partial_\mu\partial_\nu h - \Delta h_{\mu\nu})\\
&+\beta\left(\partial_\mu\partial_\nu h^{\mu\nu} - \Delta h\right) \eta_{\mu\nu}\ .
\end{split}
\end{equation} 
By rescaling the flow parameter $\lambda$ to some new flow parameter $\tau$ such that $\lambda\equiv\tau\cdot\rho$, we can impose
\begin{equation}\label{rhodef}
\alpha\cdot\rho=2\Longrightarrow\rho\equiv\frac{2}{\alpha}
\end{equation}
and accordingly define $2\vartheta\equiv\beta\cdot\rho$. Hence, the analysis contained in \cite{Headrick:2006ti} allows us to state that the well-posedness of the flow corresponds to $\alpha\geq 0$ and:
\begin{equation} 
2\vartheta\leq\frac{1}{D-1}\Longrightarrow \frac{2\beta}{\alpha}\leq\frac{1}{D-1}\ .
\end{equation} 
Namely, we are left with the simple relation
\begin{equation}\label{condbeta}
\beta\leq\frac{\alpha}{2\left(D-1\right)}\ ,
\end{equation}
with positive $\alpha$. The extremal case in which the above $\leq$ is precisely an equality corresponds to the diffusion constant of the Weyl part of the linearized equation being set to zero. From this point on, keeping \eqref{condbeta} in mind, we will simply work with the value of $\rho$ expressed in \eqref{rhodef}, so that our Ricci-Bourguignon flow will take the canonical form:
\begin{equation} \label{rbflow}
\frac{\de g_{\mu\nu}}{\de\lambda}=-2 R_{\mu\nu}+2\vartheta R g_{\mu\nu}+\nabla_{(\mu}\xi_{\nu)}\ .
\end{equation}
It is interesting to observe that, as we move towards $D\rightarrow\infty$, the $\vartheta$ parameter gets confined to being less or equal to zero. The behaviour of Swampland conjectures at a large number of space-time dimensions was recently explored in \cite{de2022geometric} and \cite{bonnefoy2021swampland}.
\subsection{Lorentzian signature}
The flow equation presented in \eqref{rbflow}, while being well-behaved and parabolic if applied to a metric with Euclidean signature, might create pathologies when considering a generic pseudo-Riemannian geometry. Nevertheless, the obvious phenomenological interest of Lorentzian metrics can justify dealing with such problems with the appropriate amount of care. In the end, we are interested in constructing (partially) on-shell flow equations for solutions of general relativity coupled to matter fields. Euclidean gravity, albeit being extremely interesting on its own and being often useful as an intermediate step towards studying standard general relativistic problems, is not -according to the current knowledge- the correct theory to make contact with observations \cite{BarberoG:1995tgc}. \\
The main issue arising when applying geometric flow equations of the form \eqref{rbflow} to a metric with signature $\left(-,+,\dots,+\right)$\footnote{Here, we have chosen to work with the typical signature assumed in literature concerning general relativity and cosmology. The discussion is unaltered for the opposite convention, namely the one which is most common in particle physics.} is that modes with timelike momentum can make the flow infinitely unstable \cite{Headrick:2006ti}. A similar issue arises when considering a simpler, diffusive flow equation for a scalar field in Minkowski spacetime. We will consider such a setting in low dimensions, as a toy model for what happens to more complicated flow equations such as \eqref{rbflow}. Therefore, let $\mathbf{M}_2$ be flat Minkowski spacetime in $D=2$ and let
\begin{equation} 
\varphi:\mathbf{M}_2\times\left[0,\lambda_{\text{max}}\right)\longrightarrow\R
\end{equation} 
be a one-parameter family of real scalar fields $\varphi_\lambda\left(t,x\right)$ on $\mathbf{M}_2$, where $\lambda_{\text{max}}$ can either be finite o infinite. After having fixed some initial configuration $\varphi_0\left(t,x\right)$, we impose the $\lambda$-behaviour of the family to be the one described by the equation:
\begin{equation} \label{heatmin}
\frac{\de\varphi_\lambda}{\de\lambda}=\Delta\varphi_\lambda=\frac{\de^2\varphi_\lambda}{\de x^2}-\frac{\de^2\varphi_\lambda}{\de t^2}\ .
\end{equation} 
The minus sign in front of the second derivative in time corresponds to a negative diffusion constant. Partial differential equations with such a feature are known to be troublesome, either due to their solution being unbounded in finite time scales or not existing at all, since they would correspond to some kind of reversed diffusion. It must be stressed that, with a generic initial condition, the problem can't be solved by simply switching the direction of the geometric flow by introducing a new flow parameter $\tau\equiv-\lambda$, since this would change both signs on the right hand side. The obstacle really lies in the mixed nature of the metric signature. The first, obvious, way out would be only to consider $t$-independent initial configurations, reducing \eqref{heatmin} to a Euclidean $2$-dimensional diffusion problem. Moving back to our original geometric flow example \eqref{rbflow} this will translate, for the purpose of the next sections, into only considering static initial metric configurations. Clearly, those are also the ones that better connect to Euclidean gravity solutions via Wick rotations. For a more detailed analysis of the subject, together with a more advanced discussion on which classes of Lorentzian manifolds admit a well-defined Ricci flow, the reader is encouraged to explore the references \cite{garcia2014homogeneity,Dhumuntarao:2018jup,Woolgar:2007vz}. From now on, we will assume to work with initial data such that the problems coming from working in Lorentzian signature can be kept under control.

\section{On-shell flow}\label{ossec}
In the following section, we address the problem of constructing on-shell flow equations for the theory presented in section \ref{theory}. As was briefly mentioned back then, the idea of providing flow equations that both keep the matter fields and the metric on shell might be too ambitious. Furthermore, the Swampland distance conjecture, as specified in the context of Ricci flow \cite{Kehagias:2019akr}, suggests that light species might enter the low energy spectrum along the flow trajectory, inevitably altering the expression for the energy-momentum tensor. Therefore, precisely looking for the explicit flow behaviour of the matter fields might not be the best available option. The most natural choice, in this sense, is to relax the on-shell condition to the metric tensor alone. In practice, this will translate into imposing some specific flow equations to the metric tensor and deriving, by forcing the equations of motion \eqref{eom} to be satisfied at every value of the flow parameter, a set of induced flow equations for the energy-momentum tensor, considered as a generic tensor on its own and without explicitly constructing it in terms of matter fields. Such problem will be postponed to section \ref{scaler}. Here, we will instead focus on deriving the general flow equations for the energy-momentum tensor and characterising their most remarkable properties when the metric is taken to follow Ricci-Bourguignon flow \eqref{rbflow}.
\subsection{General flow equations}
Given the action \eqref{action} for a theory with a dynamical spacetime metric, a cosmological constant and a set of matter fields, we can derive the equations of motion for the metric to be those expressed in \eqref{eom}. Now, we introduce a real flow parameter $\lambda\in\left[0,\lambda_{\text{max}}\right)$, where $\lambda_{\text{max}}$ is positive and can either be finite or infinite, depending on whether the flow ends up hitting a singularity. At this point, we introduce a one-parameter family of metric tensors $g_{\mu\nu}\left(\lambda\right)$, a one-parameter family of cosmological constants $\Lambda\left(\lambda\right)$ and a one-parameter family of energy-momentum tensors $T_{\mu\nu}\left(\lambda\right)$, so that the initial conditions $g_{\mu\nu}\left(0\right)$, $\Lambda\left(0\right)$ and $T_{\mu\nu}\left(0\right)$ satisfy the equations of motion \eqref{eom} and the $\lambda$-evolutions of the families are induced by the flow equations:
\be \label{floweq}
        \frac{\de g_{\mu\nu}}{\de\lambda}=A_{\mu\nu}\ ,\quad \frac{\de\Lambda}{\de\lambda}=C\ ,\quad \frac{\de T_{\mu\nu}}{\de\lambda}=\frac{1}{\kappa}B_{\mu\nu}\ .
\ee 
In the last equation of \eqref{floweq}, the $\kappa^{-1}$ factor is introduced for the sake of simplicity. Moreover, the flow-sources $A_{\mu\nu}$ and $B_{\mu\nu}$ are taken to be symmetric, $\lambda$-dependent tensors. Having defined such general flow equations, we can now move to constraining $B_{\mu\nu}$ in terms of $A_{\mu\nu}$ and $C$ in a way that preserves the equations of motion \eqref{eom} for $g_{\mu\nu}$ at any value of the flow parameter $\lambda$.
\subsection{On-shell conditions}
In order for the metric equations of motion to be conserved along the flow, together with taking an on-shell initial condition, we must impose
\be\label{condition}
\frac{\de}{\de\lambda}\left(G_{\mu\nu}+g_{\mu\nu}\Lambda\right)=\kappa\frac{\de T_{\mu\nu}}{\de\lambda}
\ee
for any value of the flow parameter $\lambda$. In order for \eqref{condition} to be a bit more concrete, we now compute the $\lambda$-evolution of the Einstein tensor $G_{\mu\nu}$ induced by \eqref{floweq}. First of all, we simply observe that we must have
\be 
\frac{\de g^{\mu\nu}}{\de\lambda}=-A^{\mu\nu}\ ,
\ee
so that the condition $g^{\mu\nu}g_{\nu\alpha}=\delta^\mu{}_\alpha$ is preserved along the flow. Furthermore, one can straightforwardly derive \cite{catino2017ricci} the flow equation for the Ricci tensor to be
\be\label{rtflow}
\begin{split}
  2\frac{\de R_{\mu\nu}}{\de\lambda}&=\nabla^\sigma\nabla_\nu A_{\mu\sigma}+\nabla_\mu\nabla^\sigma A_{\sigma\nu} -\nabla_\mu\nabla_\nu A\\
  &\quad-\Delta A_{\mu\nu }+R_\mu{}^\sigma A_{\sigma\nu }-R_\mu{}^\sigma{}_\nu {}^\theta A_{\sigma\theta}\ ,  
\end{split}
\ee
where we have defined $A\equiv g^{\mu\nu}A_{\mu\nu}$. From \eqref{rtflow}, one can obtain the following:
\be\label{rsflow}
\frac{\partial R}{\partial \lambda}=\nabla^\sigma\nabla^\theta A_{\sigma\theta}-\Delta A-A^{\sigma\theta}R_{\sigma\theta}\ .
\ee
By combining \eqref{rtflow}, \eqref{rsflow} and the flow equation for the metric, we can finally derive the flow equation for the Einstein tensor:
\be\label{etflow}
\begin{split}
    2\frac{\de G_{\mu\nu}}{\de\lambda}&=\nabla^\sigma\nabla_\nu A_{\mu\sigma}+\nabla_\mu\nabla^\sigma A_{\sigma\nu} -\nabla_\mu\nabla_\nu A-\Delta A_{\mu\nu }\\
  &\quad 
  -g_{\mu\nu}\nabla^\sigma\nabla^\theta A_{\sigma\theta}+g_{\mu\nu}\Delta A+g_{\mu\nu}A^{\sigma\theta}R_{\sigma\theta}\\
    &\quad+R_\mu{}^\sigma A_{\sigma\nu }-R_\mu{}^\sigma{}_\nu {}^\theta A_{\sigma\theta}-A_{\mu\nu}R\ .
\end{split}
\ee
With the expression derived in \eqref{etflow}, together with the flow equations for the cosmological constant and the energy-momentum tensor, we finally obtain the on-shell condition for the metric:
\be\label{onshellB}
\begin{split}
   2B_{\mu\nu}&=\nabla^\sigma\nabla_\nu A_{\mu\sigma}+\nabla_\mu\nabla^\sigma A_{\sigma\nu} -\nabla_\mu\nabla_\nu A-R_\mu{}^\sigma{}_\nu {}^\theta A_{\sigma\theta}\\
       &\quad+2C g_{\mu\nu}-\Delta A_{\mu\nu }+R_\mu{}^\sigma A_{\sigma\nu }-A_{\mu\nu}R+2\Lambda A_{\mu\nu}\\
  &\quad 
  -g_{\mu\nu}\nabla^\sigma\nabla^\theta A_{\sigma\theta}+g_{\mu\nu}\Delta A+g_{\mu\nu}A^{\sigma\theta}R_{\sigma\theta}\ .
\end{split}
\ee
Therefore, by freely choosing $A_{\mu\nu}$ and $C$ and imposing $B_{\mu\nu}$ to precisely be defined by \eqref{onshellB}, we are guaranteed that the overall flow conserves the equations of motion \eqref{eom} for the metric tensor. Once more, it is crucial to stress that this has nothing to do with the specific matter content of the theory being on-shell. We are generically treating the energy-momentum tensor as an appropriate source for space-time curvature. Assessing whether particular theories, with their own matter contents and the corresponding on-shell conditions to be solved together with \eqref{onshellB}, allow for the full set of equations of motion to be conserved along the $\lambda$-evolution is a completely different and way harder problem.
\subsection{Ricci-Bourguignon flow}
In the following discussion, we will consider the specific case in which the flow source $A_{\mu\nu}$ for the metric is precisely the one expressed in \eqref{rbflow}, neglecting the diffeomorphism part.
Hence, by taking
\be
A_{\mu\nu}=-2R_{\mu\nu}+2\vartheta Rg_{\mu\nu}
\ee
we can derive $A=2\left(D\vartheta-1\right)R$ and the following relation between $C$ and $B_{\mu\nu}$:
\be\label{rbosB}
\begin{split}
   B_{\mu\nu}&=-\nabla^\sigma\nabla_\nu R_{\mu\sigma}+\left[\left(2-D\right)\vartheta+\frac{1}{2}\right]\nabla_\mu\nabla_\nu R\\
   &\quad
   +R_\mu{}^\sigma{}_\nu {}^\theta R_{\sigma\theta}+C g_{\mu\nu}+\Delta R_{\mu\nu }-R_\mu{}^\sigma R_{\sigma\nu }\\
       &\quad+R_{\mu\nu}R-g_{\mu\nu}\left[\left(2-D\right)\vartheta +\frac{1}{2}\right]\Delta R\\
  &\quad-2\Lambda R_{\mu\nu}+2\vartheta\Lambda Rg_{\mu\nu}-g_{\mu\nu}R^{\sigma\theta}R_{\sigma\theta}\ .
\end{split}
\ee
This way, we have explicitly derived a relation for $B_{\mu\nu}$ and $C$, in terms of $\vartheta$, for the case in which the metric follows Ricci-Bourguignon flow. At this point, we want to investigate the linear approximation of the flow equation for $T_{\mu\nu}$, sourced by $B_{\mu\nu}$. In order to do so, we consider small and $\lambda$-dependent perturbations of the metric
\be 
g_{\mu\nu}\left(\lambda\right)\approx\eta_{\mu\nu}+\epsilon h_{\mu\nu}\left(\lambda\right)\ ,
\ee
with $|\epsilon|\ll 1$, and linearise $B_{\mu\nu}$ in order to only keep terms up to the first power in $\varepsilon$. In this sense, we must once more use the formulas \eqref{fort} and \eqref{fors}. Nevertheless, it is not necessary to plug them directly into \eqref{rbosB}. By starting from the equations on motion \eqref{eom} and linearising them, we can derive the needed result in a much faster way. In that sense, we expand the energy-momentum too in $\epsilon$ around the vacuum as:
\be
T_{\mu\nu}\left(\lambda\right)\approx\epsilon t_{\mu\nu}\left(\lambda\right)+\mathcal{O}\left(\epsilon^2\right)\ .
\ee
Hence, at a linear level, the equations of motion get to be:
\be\label{line}
\partial_\sigma\partial_\mu h^\sigma_\nu + \partial_\sigma\partial_\nu h^\sigma_\mu - \partial_\mu\partial_\nu h - \de^2 h_{\mu\nu} - \eta_{\mu\nu}\partial_\rho\partial_\lambda h^{\rho\lambda} + \eta_{\mu\nu}\de^2 h=2\kappa t_{\mu\nu}\ .
\ee
At this point, we can take $\lambda$ derivatives on both sides. This is made simple by the fact that they commute with spacetime derivatives. Taking $b_{\mu\nu}$ and $a_{\mu\nu}$ to be the linear-order expansions of $B_{\mu\nu}$ and $A_{\mu\nu}$, respectively, we then get:
\be\label{linefl}
2b_{\mu\nu}=\partial_\sigma\partial_\mu a^\sigma_\nu + \partial_\sigma\partial_\nu a^\sigma_\mu - \partial_\mu\partial_\nu a - \de^2 a_{\mu\nu} - \eta_{\mu\nu}\partial_\rho\partial_\lambda a^{\rho\lambda} + \eta_{\mu\nu}\de^2 a\ .
\ee
Now, we must derive an explicit expression for $a_{\mu\nu}$ from \eqref{rbflow}. We obtain:
\be\label{a}
a_{\mu\nu}=2\vartheta \left(\partial_\alpha\partial_\beta h^{\alpha\beta} - \Delta h\right) \eta_{\mu\nu}-(\partial_\tau\partial_\mu h^\tau_\nu + \partial_\tau\partial_\nu h^\tau_\mu - \partial_\mu\partial_\nu h - \de^2 h_{\mu\nu})\ .
\ee
Furthermore, for the trace we have
\be\label{at}
a=\eta^{\mu\nu}a_{\mu\nu}=2\left(D\vartheta -1\right)\left(\partial_\alpha\partial_\beta h^{\alpha\beta} - \Delta h\right) \ ,
\ee 
which is consistent with directly linearising the explicit expression for $A$. By plugging \eqref{a} and \eqref{at} into \eqref{linefl}, we are left with:
\be 
\begin{split}
    2b_{\mu\nu}&=2\vartheta\left(2-D\right) \partial_\nu\partial_\mu\left(\partial_\alpha\partial_\beta h^{\alpha\beta} - \de^2 h\right) \\
    &\quad+\de^2 (\partial_\tau\partial_\mu h^\tau_\nu + \partial_\tau\partial_\nu h^\tau_\mu - \partial_\mu\partial_\nu h - \de^2 h_{\mu\nu})\\
    &\quad+ \left[2\vartheta\left(D-2\right) -1 \right]\de^2 \left(\partial_\alpha\partial_\beta h^{\alpha\beta} - \de^2 h\right) \eta_{\mu\nu} \ .
\end{split}
\ee
It can be shown that our condition on $\vartheta$ consistently applies here. Exploiting the harmonic gauge, in which 
\be
\de^\mu h_{\mu\nu}=\frac{1}{2}\de_\nu h\ ,
\ee
the expression for $b_{\mu\nu}$ simplifies to:
\be
\begin{split}
    2b_{\mu\nu}&=\vartheta\left(D-2\right)\partial^2\left( \partial_\nu\partial_\mu-\eta_{\mu\nu}\de^2 \right) h \\
    &\quad+\de^2\de^2 \left(\frac{h}{2} \eta_{\mu\nu}-h_{\mu\nu}\right)\ .
\end{split}
\ee
By defining the symmetric tensor
\be
\Bar{h}_{\mu\nu}=h_{\mu\nu}-\frac{h}{2} \eta_{\mu\nu}\ ,
\ee
we get $2\Bar{h}=\left(2-D\right)h$ and:
\be
\begin{split}
    b_{\mu\nu}&=\vartheta\partial^2\left(\eta_{\mu\nu}\de^2-\partial_\nu\partial_\mu\right) \Bar{h} -\frac{1}{2}\de^2\de^2\Bar{h}_{\mu\nu}\ .
\end{split}
\ee
Therefore, the final form is rather simple to understand and analyse and gives the following flow equation for the first-order contribution in the $\epsilon$-expansion of $T_{\mu\nu}$:
\be 
\frac{\de t_{\mu\nu}}{\de\lambda}=\frac{\vartheta}{\kappa}\partial^2\left(\eta_{\mu\nu}\de^2-\partial_\nu\partial_\mu\right) \Bar{h} -\frac{1}{2\kappa}\de^2\de^2\Bar{h}_{\mu\nu}\ .
\ee 
The flow is naturally well-defined, since it directly comes from the well-defined flow of $g_{\mu\nu}$ and the resulting flow of the Einstein tensor $G_{\mu\nu}$.
\subsubsection{Einstein manifolds}
In the following discussion, we consider the particular case in which the metric initial condition is an Einstein manifold. Indeed, we take $g_{\mu\nu}\left(0\right)$ so that:
\be\label{einst}
R_{\mu\nu}=\frac{R}{D}g_{\mu\nu}\ .
\ee
Concerning the Riemann tensor, we have:
\be 
R_{\alpha\mu\beta\nu}=\frac{R}{D(D-1)}\left(g_{\alpha\beta}g_{\mu\nu}-g_{\alpha\nu}g_{\mu\beta}\right)\ .
\ee
In order for such an initial condition to be on-shell, we must take
\be 
T_{\mu\nu}=Kg_{\mu\nu}
\ee
and solve \eqref{eom} as:
\begin{equation}
   \left(\frac{R}{D}-\frac{R}{2}+\Lambda-K\right)g_{\mu\nu}=0\ .
\end{equation}
Namely, we have:
\be\label{curv}
R=\left(K-\Lambda\right)\frac{2D}{2-D}\ .
\ee
In can be easily shown \cite{DeBiasio:2020xkv,catino2017ricci} that a flow of the form presented in \eqref{rbflow} conserves the property \eqref{einst} of the initial condition. Namely, that an Einstein manifold remains an Einstein manifold under Ricci-Bourguignon flow. In particular, the flow equation for $R$, as derived from \eqref{rsflow} is: 
\be
\frac{\partial R}{\partial \lambda}=\frac{2\left(1-\vartheta\right)}{D}R^2\ .
\ee
The above can be simply solved as:
\be
R\left(\lambda\right)=\frac{DR_0}{D-2\left(1-\vartheta\right)R_0\lambda}\ .
\ee
Furthermore, given the flow equation \eqref{floweq} for $\Lambda$, we can explicitly write its behaviour:
\be 
\Lambda\left(\lambda\right)=\Lambda_0+\int^{\lambda}_0 C\left(\tau\right)\diff\tau\ .
\ee
Now, we can safely plug \eqref{einst} into \eqref{onshellB}. We are left with:
\be
\begin{split}
   2B_{\mu\nu}&=\left[\left(2-D\right)\vartheta+\frac{D-2}{2D}\right]\nabla_\mu\nabla_\nu R\\
       &\quad-g_{\mu\nu}\left[\left(2-D\right)\vartheta +\frac{D-2}{2D}\right]\Delta R\\
  &\quad+2\Lambda \frac{D\vartheta-1}{D}Rg_{\mu\nu}+C g_{\mu\nu} \ .
\end{split}
\ee
The assumption of constant curvature directly implies:
\be
\begin{split}
   B_{\mu\nu}&=\left(2\Lambda \frac{D\vartheta-1}{D}R+C \right)g_{\mu\nu} \ .
\end{split}
\ee
Anyway, we can once more read-off the flow behaviour of the energy momentum tensor directly from \eqref{curv}, which gives us:
\be 
K\left(\lambda\right)=\frac{R_0\left(2-D\right)}{2D-4\left(1-\vartheta\right)R_0\lambda}+\Lambda_0+\int^{\lambda}_0 C\left(\tau\right)\diff\tau\ .
\ee 
Therefore, starting with an on-shell Einstein manifold, imposing the metric to evolve according to \eqref{rbflow} and freely choosing the flow behaviour of the cosmological constant, we are left with the explicit flow behaviour for the constant $K$ appearing in the energy-momentum tensor. 

\section{On-shell flow and entropy functional}\label{entropysection}
In this section, we consider a general theory of the form outlined in section \ref{theory}. Furthermore, we introduce an entropy functional $\mathcal{F}\left(g_{\mu\nu},\varphi_1,\dots,\varphi_N\right)$, depending on both the metric and the $N$ matter fields, so that the flow equations can be derived as the gradient flow equations
\be 
\frac{\de g_{\mu\nu}}{\de\lambda}=-\frac{\delta\mathcal{F}}{\delta g^{\mu\nu}}\ ,\quad \frac{\de \varphi_A}{\de\lambda}=-\frac{\delta\mathcal{F}}{\delta\varphi_A}\ ,
\ee 
where the variations with respect to the fields have to be performed in an appropriate, volume-preserving way \cite{Kehagias:2019akr,Perelman:2006un,de2022geometric}. This way, by plugging the above implicit expression into \eqref{onshellB}, we have that the on-shell condition for the metric can be achieved by imposing the energy-momentum tensor to evolve according to the flow-source:
\be\label{Bentropy}
\begin{split}
   2B_{\mu\nu}&=-\nabla^\sigma\nabla_\nu \frac{\delta\mathcal{F}}{\delta g^{\mu\sigma}}-\nabla_\mu\nabla^\sigma \frac{\delta\mathcal{F}}{\delta g^{\sigma\nu}} +g^{\alpha\beta}\nabla_\mu\nabla_\nu \frac{\delta\mathcal{F}}{\delta g^{\alpha\beta}}+R_\mu{}^\sigma{}_\nu {}^\theta \frac{\delta\mathcal{F}}{\delta g^{\sigma\theta}}\\
       &\quad-2C g_{\mu\nu}+\Delta \frac{\delta\mathcal{F}}{\delta g^{\mu\nu}}-R_\mu{}^\sigma \frac{\delta\mathcal{F}}{\delta g^{\sigma\nu}}+R\frac{\delta\mathcal{F}}{\delta g^{\mu\nu}}-2\Lambda \frac{\delta\mathcal{F}}{\delta g^{\mu\nu}}\\
  &\quad 
  +g_{\mu\nu}\nabla^\sigma\nabla^\theta \frac{\delta\mathcal{F}}{\delta g^{\sigma\theta}}-g_{\mu\nu}g^{\alpha\beta}\Delta \frac{\delta\mathcal{F}}{\delta g^{\alpha\beta}}-g_{\mu\nu}R^{\sigma\theta}\frac{\delta\mathcal{F}}{\delta g^{\sigma\theta}}\ .
\end{split}
\ee
Nevertheless, it must be observed that the $\mathcal{F}$-entropy functional itself induces a flow 
\be 
\frac{\de T_{\mu\nu}}{\de\lambda}=E_{\mu\nu}
\ee 
of the energy-momentum tensor, where the flow source $E_{\mu\nu}$ is given by:
\be 
E_{\mu\nu}=-\frac{\delta T_{\mu\nu}}{\delta g_{\alpha\beta}}\frac{\delta\mathcal{F}}{\delta g^{\alpha\beta}}-\sum_{A=1}^N\frac{\delta T_{\mu\nu}}{\delta\varphi_A}\frac{\delta\mathcal{F}}{\delta\varphi_A}\ .
\ee 
Here a crucial issue that requires particular care arises. In fact, one has no reason to believe the above two flow-sources for the energy-momentum tensor match. In general, we have:
\be 
S_{\mu\nu}\equiv E_{\mu\nu}-B_{\mu\nu}\neq 0\ .
\ee 
Therefore, if we both want our metric tensor to be on-shell\footnote{Similar statements can be made on the matter fields, making the problem even more complicated.} and our flow to be realised as the gradient flow of $\mathcal{F}$, there are two directions we might choose to take. The first option is to look for a set of matter fields and an appropriate entropy functional $\mathcal{F}$ so that the metric, and perhaps also the matter, stays on-shell along the corresponding gradient flow. This would deal with the problem at its root. Unfortunately, the definition of $\mathcal{F}$ -if possible- would be an extremely complicated task. This is the reason for which we will set this possibility apart and focus on the other one. The second option is to assume the difference between $B_{\mu\nu}$ and $E_{\mu\nu}$ to be accounted for by the appearance of further matter fields. Hence, we could write such a difference down and try to read them off. This intuition is obviously motivated by the Swampland distance conjecture \cite{Ooguri:2006in}, that claims we should expect towers of light states to emerge for large fields displacements in the moduli space. In order to follow the outlined procedure, one must introduce the symmetric tensor
\be
S_{\mu\nu}\equiv B_{\mu\nu}-E_{\mu\nu}
\ee
and derive its explicit form. Once $S_{\mu\nu}$ is derived, one can write the energy momentum tensor as
\be 
T_{\mu\nu}\left(\lambda\right)\equiv \Bar{T}_{\mu\nu}\left(\lambda\right)+\Hat{T}_{\mu\nu}\left(\lambda\right)\ ,
\ee 
with $\Hat{T}_{\mu\nu}\left(0\right)=0$ and:
\be 
\frac{\de \Bar{T}_{\mu\nu}}{\de\lambda}=E_{\mu\nu}\ ,\quad\frac{\de \Hat{T}_{\mu\nu}}{\de\lambda}=S_{\mu\nu}\ .
\ee
This way, we have an energy-momentum tensor component $\Bar{T}_{\mu\nu}$ produced by the scalar $\phi$ and evolving according to the appropriate gradient-flow equations, together with an extra $\Hat{T}_{\mu\nu}$ term appearing along the flow accounting for the emergence of new matter fields, which can flow themselves, so that the metric remains on-shell. This procedure can be naturally generalised to any initial matter content. The extension to the case in which one imposes the matter field to be too on-shell is straightforward and natural.
\section{Scalar fields in curved space-time}\label{scaler}
In the following section, we move from treating the energy-momentum tensor as a generic symmetric tensorial source for Einstein field equations and try to characterise it in terms of a specific set of fields. Practically, we choose to work with a theory of the form presented in section \ref{theory} and assume its matter content to be comprised of $N$ real scalar fields $\varphi_i$. Therefore, we have that the associated canonically normalised Lagrangian density\footnote{For the signature choice $\left(-,+,\dots,+\right)$.} is:
\begin{equation}
    \mathcal{L}_M=-\sum_{i=1}^N\frac{1}{2}\nabla^\mu\varphi_i\nabla_\mu\varphi_i-\sum_{i,j=1}^N\gamma_{ij}\varphi_i\varphi_j\ .
\end{equation}
The precise form of the energy-momentum tensor can be shown to be:
\begin{equation}
    \begin{split}
        T_{\mu\nu}&=\sum_{i=1}^N\nabla_\mu\varphi_i\nabla_\nu\varphi_i-g_{\mu\nu}\left(\sum_{i=1}^N\frac{1}{2}\nabla^\alpha\varphi_i\nabla_\alpha\varphi_i+\sum_{i,j=1}^N\gamma_{ij}\varphi_i\varphi_j\right)\ .
    \end{split}
\end{equation}
At this point, considering the flow equations presented in \eqref{floweq}, we assume to work with the metric flow source:
\be 
\begin{split}\label{mss}
     A_{\mu\nu}&\equiv -2R_{\mu\nu}+2\vartheta Rg_{\mu\nu}+\sum_{j=1}^N b_j\nabla_\mu\nabla_\nu\varphi_j+g_{\mu\nu}\sum_{j=1}^N d_j\Delta\varphi_j\\
     &\quad+\sum_{i,j=1}^N e_{ij}\nabla_\mu\varphi_i\nabla_\nu\varphi_j+g_{\mu\nu}\sum_{i,j=1}^N f_{ij}\nabla_\alpha\varphi_i\nabla^\alpha\varphi_j\ .
\end{split}
\ee 
Its trace is simply:
\begin{equation}
    A=2\left(D\vartheta-1\right)R+\sum_{j=1}^N\left(Dd_j+b_j\right)\Delta\varphi_j+\sum_{i,j=1}^N\left(Df_{ij}+e_{ij}\right)\nabla_\alpha\varphi_i\nabla^\alpha\varphi_j\ .
\end{equation}
At this point, by leaving the flow source $C$ for the cosmological constant implicit, we can derive the explicit for $B_{\mu\nu}$ by plugging everything into the expression \eqref{onshellB}
\be 
\begin{split}\label{souceonesc}
    2B_{\mu\nu}&=-2\nabla^\sigma\nabla_\nu R_{\mu\sigma}+2\left[1+\left(2-D\right)\vartheta\right]\nabla_\mu\nabla_\nu R+2\Delta R_{\mu\nu}\\
&\quad-2\nabla_\mu\nabla^\sigma R_{\sigma\nu}+\left[2\left(D-2\right)\vartheta-1 \right]g_{\mu\nu}\Delta R\\
    &\quad+2R_\mu{}^\sigma{}_\nu {}^\theta R_{\sigma\theta}+2RR_{\mu\nu}-2R_\mu{}^\sigma R_{\sigma\nu}-4\Lambda R_{\mu\nu}\\
    &\quad+2\left(C-R_{\sigma\theta}R^{\sigma\theta}+2\Lambda\vartheta R\right)g_{\mu\nu}+\sum_{j=1}^N\Omega^{\ j}_{\mu\nu}\varphi_j\\
        &\quad     +\sum_{i,j=1}^N\Theta_{\mu\nu}^{\ ij}\nabla_\alpha\varphi_i\nabla^\alpha\varphi_j+\sum_{i,j=1}^N\Sigma_{\mu\nu}^{\sigma\theta ij}\nabla_\sigma\varphi_i\nabla_\theta\varphi_j\ .
\end{split}
\ee
Where we have introduced $N$ metric-dependent differential operators
\be 
\begin{split}
    \Omega^{\ j}_{\mu\nu}&= b_j\nabla_\mu\Delta\nabla_\nu-\left[\left(D-2\right)d_j+b_j\right]\nabla_\mu\nabla_\nu\Delta- b_j\Delta\nabla_\mu\nabla_\nu\\
    &\quad+ \left[b_j\left(2\Lambda\delta_\mu{}^\sigma\delta_\nu{}^{\theta}+g_{\mu\nu}R^{\sigma\theta}\right)\nabla_\sigma\nabla_\theta\right]+ b_j\nabla^\sigma\nabla_\nu \nabla_\mu\nabla_\sigma\\
    &\quad+g_{\mu\nu}\left\{\left[\left(D-2\right)d_j+b_j\right]\Delta\Delta-b_j\nabla^\sigma\nabla^\theta\nabla_\sigma\nabla_\theta+2d_j\Lambda\Delta\right\}\\
    &\quad+\left[\left(R_\mu{}^\sigma\delta_\nu{}^{\theta}-R_\mu{}^\sigma{}_\nu {}^\theta-R\delta_\mu{}^\sigma\delta_\nu{}^{\theta}\right) b_j\nabla_\sigma\nabla_\theta\right]
\end{split}
\ee
and $2N^2$ metric-dependent differential operators:
\be
\Theta_{\mu\nu}^{\ ij}\equiv\left(g_{\mu\nu}\Delta-\nabla_\mu\nabla_\nu \right)\left(Df_{ij}+e_{ij}\right)+2\left(\nabla_\mu\nabla_\nu-g_{\mu\nu}\Delta+\Lambda g_{\mu\nu}\right)f_{ij}\ ,
\ee
\be
\begin{split}
   \Sigma_{\mu\nu}^{\sigma\theta ij}&\equiv \Big[\left(R_\mu{}^\sigma+\nabla_\mu\nabla^\sigma \right)\delta_\nu^{\ \theta}+g_{\mu\nu}R^{\sigma\theta}-g_{\mu\nu}\nabla^\sigma\nabla^\theta -R_\mu{}^\sigma{}_\nu {}^\theta\\
&\quad+\left(2\Lambda-R-\Delta \right)\delta_{\mu}^{\ \sigma}\delta_{\nu}^{\ \theta}+\delta_{\mu}^{\ \sigma}\nabla^\theta\nabla_\nu\Big]e_{ij} \ .
\end{split}
\ee
The details of the above computation can be found in appendix \ref{compute}.
\subsection{Single free scalar field}\label{singlesc}
By assuming to work with a single scalar field $\phi$, the metric flow source reduces to:
\be\label{singsource}
\begin{split}
         A_{\mu\nu}&\equiv -2R_{\mu\nu}+2\vartheta Rg_{\mu\nu}+ b\nabla_\mu\nabla_\nu\phi+g_{\mu\nu}d\Delta\phi\\
     &\quad+e\nabla_\mu\phi\nabla_\nu\phi+fg_{\mu\nu}\nabla_\alpha\phi\nabla^\alpha\phi\ .
\end{split}
\ee
Consequently, the energy-momentum tensor flow source gets to be
\be\label{osbssi}
\begin{split}
   2B_{\mu\nu}&=-2\nabla^\sigma\nabla_\nu R_{\mu\sigma}+2\left[1+\left(2-D\right)\vartheta\right]\nabla_\mu\nabla_\nu R+2\Delta R_{\mu\nu}\\
&\quad-2\nabla_\mu\nabla^\sigma R_{\sigma\nu}+\left[2\left(D-2\right)\vartheta-1 \right]g_{\mu\nu}\Delta R\\
    &\quad+2R_\mu{}^\sigma{}_\nu {}^\theta R_{\sigma\theta}+2RR_{\mu\nu}-2R_\mu{}^\sigma R_{\sigma\nu}-4\Lambda R_{\mu\nu}\\
    &\quad+2\left(C-R_{\sigma\theta}R^{\sigma\theta}+2\Lambda\vartheta R\right)g_{\mu\nu}+\Omega_{\mu\nu}\phi\\
    &\quad     +\Theta_{\mu\nu}\nabla_\alpha\phi\nabla^\alpha\phi+\Sigma_{\mu\nu}^{\sigma\theta}\nabla_\sigma\phi\nabla_\theta\phi\ .
\end{split}
\ee
Given the analysis performed in \cite{DeBiasio:2022omq}, we now want to find an entropy $\mathcal{F}$-functional
\begin{equation}\label{funk}
        \mathcal{F}_{\left(\alpha,\gamma\right)}\left(g,\phi\right)=\int\diff^{D}x\sqrt{g}e^{-\phi}\left[R+\alpha\left(\nabla\phi\right)^2+\gamma\Delta\phi\right]
\end{equation}
such that the flow sourced by \eqref{singsource} can be derived as its associated gradient flow equation for $g_{\mu\nu}$ and $\phi$. As discussed in the above-mentioned reference, we have that the volume-preserving gradient of \eqref{funk} gives:
\begin{equation}\label{geneqm}
    \begin{split}
        \frac{\de g_{\mu\nu}}{\de\lambda}=&-2R_{\mu\nu}+2\left(1-\alpha+\gamma\right)\nabla_{\mu}\phi\nabla_{\nu}\phi-2\left(1+2\gamma\right)\nabla_{\nu}\nabla_{\mu}\phi+\\
        &+2\left(1-\alpha-\frac{\gamma}{2}\right)g_{\mu\nu}\Delta\phi-2\left(1-\alpha-\frac{\gamma}{2}\right)g_{\mu\nu}\left(\nabla\phi\right)^2\ ,
            \end{split}
\end{equation}
\begin{equation}\label{geneqs}
    \begin{split}
        \frac{\de\phi}{\de\lambda}=&-R+\left[\left(1-\alpha\right)\left(1-D\right)+\gamma\left(1+\frac{D}{2}\right)\right]\left(\nabla\phi\right)^{2}+\\
        &-\left[1-D\left(1-\alpha\right)+\gamma\left(2+\frac{D}{2}\right)\right]\Delta\phi\ .
    \end{split}
\end{equation}
In order to match \eqref{singsource} with \eqref{geneqm}, we first have impose $\vartheta=0$. Furthermore, we must take $d=-f$ and select the following parameters:
\be 
\begin{split}
    b=-2\left(1+2\gamma\right)\ ,\quad d=-f=2\left(1-\alpha-\frac{\gamma}{2}\right)\ ,\quad e=2\left(1-\alpha+\gamma\right)\ .
\end{split}
\ee
The above equations directly imply:
\be 
\gamma=-\frac{b+2}{4}\ ,\quad \alpha=\frac{10-4d+b}{8}\ .
\ee
Hence, $e$ too gets fixed in terms of $b$ and $d$ as:
\be 
e=\frac{2+4d+b}{4}\ .
\ee 
Therefore, the restricted flow 
\be\label{mf}
\begin{split}
         \frac{\de g_{\mu\nu}}{\de\lambda}&\equiv -2R_{\mu\nu}+ b\nabla_\mu\nabla_\nu\phi+g_{\mu\nu}d\Delta\phi\\
         &\quad+\frac{2+4d+b}{4}\nabla_\mu\phi\nabla_\nu\phi-dg_{\mu\nu}\nabla_\alpha\phi\nabla^\alpha\phi
\end{split}
\ee
can be derived as the volume-preserving metric gradient flow of the entropy $\mathcal{F}$-functional:
\begin{equation}
        \mathcal{F}\left(g,\phi\right)=\int\diff^{D}x\sqrt{g}e^{-\phi}\left[R+\frac{10-4d+b}{8}\left(\nabla\phi\right)^2-\frac{b+2}{4}\Delta\phi\right]\ .
\end{equation}
If we wanted to also consider the associated scalar flow, we would have:
\begin{equation}\label{sf}
    \begin{split}
        \frac{\de\phi}{\de\lambda}=&-R+\left[\frac{4d-b-2}{8}\left(1-D\right)-\frac{b+2}{4}\left(1+\frac{D}{2}\right)\right]\left(\nabla\phi\right)^{2}+\\
        &-\left[1-D\frac{4d-b-2}{8}-\frac{b+2}{4}\left(2+\frac{D}{2}\right)\right]\Delta\phi\\
        &\equiv-R+f_D\left(\nabla\phi\right)^{2}-g_D\Delta\phi\ .
    \end{split}
\end{equation}
What would assuming \eqref{mf} and \eqref{sf} imply for the energy-momentum tensor $T_{\mu\nu}$? In order to investigate such question, we start from
\be 
T_{\mu\nu}=\nabla_\mu\phi\nabla_\nu\phi-g_{\mu\nu}\left(\frac{1}{2}\nabla_\alpha\phi\nabla^\alpha\phi+\frac{m^2}{2}\phi^2\right)\ ,
\ee 
where $m$ is the mass of the scalar field $\phi$. By taking the $\lambda$-derivative, we have:
\be
\begin{split}\label{indflowT}
    \frac{\de T_{\mu\nu}}{\de\lambda}\equiv E_{\mu\nu}&=-\nabla_\mu\phi\nabla_\nu R+f_D\nabla_\mu\phi\nabla_\nu\left(\nabla\phi\right)^{2}-g_D\nabla_\mu\phi\nabla_\nu\Delta\phi\\
    &\quad-\nabla_\nu\phi\nabla_\mu R+f_D\nabla_\nu\phi\nabla_\mu\left(\nabla\phi\right)^{2}-g_D\nabla_\nu\phi\nabla_\mu\Delta\phi\\
    &\quad-g_{\mu\nu}m^2\phi\left[-R+f_D\left(\nabla\phi\right)^{2}-g_D\Delta\phi\right]\\
    &\quad+g_{\mu\nu}\left[\nabla^\alpha\phi\nabla_\alpha R-f_D\nabla^\alpha\phi\nabla_\alpha\left(\nabla\phi\right)^{2}+g_D\nabla^\alpha\phi\nabla_\alpha\Delta\phi\right]\\
    &\quad+2R_{\mu\nu}\left(\frac{1}{2}\nabla_\alpha\phi\nabla^\alpha\phi+\frac{m^2}{2}\phi^2\right)\\
    &\quad- b\nabla_\mu\nabla_\nu\phi\left(\frac{1}{2}\nabla_\alpha\phi\nabla^\alpha\phi+\frac{m^2}{2}\phi^2\right)\\
    &\quad-g_{\mu\nu}d\Delta\phi\left(\frac{1}{2}\nabla_\alpha\phi\nabla^\alpha\phi+\frac{m^2}{2}\phi^2\right)\\
    &\quad-\frac{2+4d+b}{4}\nabla_\mu\phi\nabla_\nu\phi\left(\frac{1}{2}\nabla_\alpha\phi\nabla^\alpha\phi+\frac{m^2}{2}\phi^2\right)\\
         &\quad+dg_{\mu\nu}\nabla_\alpha\phi\nabla^\alpha\phi\left(\frac{1}{2}\nabla_\alpha\phi\nabla^\alpha\phi+\frac{m^2}{2}\phi^2\right)\ .
\end{split}
\ee
By selecting a specific flow for the metric tensor, we have thus obtained it as the gradient flow from an $\mathcal{F}$-entropy functional. Therefore, we managed to derived the associated scalar flow equation. Such a discussion directly implies an energy-momentum tensor flow sourced by \eqref{indflowT}. Conversely, by selecting the same flow for the metric tensor and imposing Einstein field equations to be on-shell along the flow, we have obtained an explicit expression \eqref{osbssi} for the required flow source of the energy-momentum tensor. Being the two different, we will apply the procedure described in section \ref{entropysection} and try to account for it by introducing extra matter fields in specific examples.
\subsubsection{An example in $D=2$}\label{specific}
We now focus on a specific and simple example, which will allow us to derive explicit and insightful results. First of all, we assume to work in $D=2$, with coordinates $\left(t,x\right)$. This way we can take a very simple ansatz for our static metric:
\be 
g_{\mu\nu}=\begin{pmatrix}
-g_1\left(x\right) & 0 \\
a & g_2\left(x\right)
\end{pmatrix}
\ee
Moreover, we assume the scalar field to be time-independent. The fact that we work in $D=2$ implies that:
\be 
R_{\mu\nu}=\frac{R}{2}g_{\mu\nu}\ .
\ee
It can be easily shown that all the above assumptions are conserved along the flow. On top of that, the non-trivial Christoffel symbols are:
\be 
\Gamma^t_{\ tx}=\Gamma^t_{\ xt}=\frac{g_1'}{2g_1}\ ,\quad\Gamma^x_{\ tt}=\frac{g_1'}{2g_2}\ ,\quad\Gamma^x_{\ xx}=\frac{g_2'}{2g_2}\ .
\ee 
We simply have $\nabla_\alpha\phi\nabla^\alpha\phi=\left(\phi'\right)^2/g_2$, where the prime-symbol stands for an $x$-derivative. In order to make our setting even simpler, we choose $d=b=0$ in the flow source $A_{\mu\nu}$, so that also $f=0$ and $e=1/2$, consistently with the derivation of our flow equations from the $\mathcal{F}$-functional. So, they reduce to:
\be\label{specflmet}
\begin{split}
         \frac{\de g_{\mu\nu}}{\de\lambda}&= -2R_{\mu\nu}+\frac{1}{2}\nabla_\mu\phi\nabla_\nu\phi\ ,
\end{split}
\ee
\begin{equation}
    \begin{split}
        \frac{\de\phi}{\de\lambda}=&-R-\frac{3}{4}\left(\nabla\phi\right)^{2}\ .
    \end{split}
\end{equation}
Before writing them explicitly in terms of $g_1$ and $g_2$, we observe that:
\be 
R=\frac{g_2 \left[\left(g_1'\right)^2-2 g_1 g_1''\right]+g_1 g_1'g_2'}{2 g_1^2 g_2^2}
\ee 
Therefore, the three equations are:
\be
\begin{split}
         \frac{\de g_1}{\de\lambda}= -\frac{g_2 \left[\left(g_1'\right)^2-2 g_1 g_1''\right]+g_1 g_1'g_2'}{2 g_1 g_2^2}=-Rg_1\ ,
\end{split}
\ee
\be
\begin{split}
         \frac{\de g_2}{\de\lambda}=\equiv -\frac{g_2 \left[\left(g_1'\right)^2-2 g_1 g_1''\right]+g_1 g_1'g_2'}{2 g_1^2 g_2}+\frac{\left(\phi'\right)^2}{2}=-Rg_2+\frac{\left(\phi'\right)^2}{2}\ ,
\end{split}
\ee
\begin{equation}
    \begin{split}
        \frac{\de\phi}{\de\lambda}=&-\frac{g_2 \left[\left(g_1'\right)^2-2 g_1 g_1''\right]+g_1 g_1'g_2'}{2 g_1^2 g_2^2}-\frac{3\left(\phi'\right)^2}{4g_2}=-R-\frac{3\left(\phi'\right)^2}{4g_2}\ .
    \end{split}
\end{equation}
Concerning the differential operators appearing in $B_{\mu\nu}$, we have $\Omega_{\mu\nu}=0$,
\be
\Theta_{\mu\nu}=\frac{1}{2}\left(g_{\mu\nu}\Delta-\nabla_\mu\nabla_\nu \right)\ ,
\ee
and, finally:
\be
\begin{split}
   \Sigma_{\mu\nu}^{\sigma\theta }&=\frac{1}{2}\Big[\left(R_\mu{}^\sigma+\nabla_\mu\nabla^\sigma \right)\delta_\nu^{\ \theta}+g_{\mu\nu}R^{\sigma\theta}-g_{\mu\nu}\nabla^\sigma\nabla^\theta -R_\mu{}^\sigma{}_\nu {}^\theta\\
&\quad+\left(2\Lambda-R-\Delta \right)\delta_{\mu}^{\ \sigma}\delta_{\nu}^{\ \theta}+\delta_{\mu}^{\ \sigma}\nabla^\theta\nabla_\nu\Big] \ .
\end{split}
\ee
Since we are in $D=2$, we simply set $\Lambda=C=0$. Thus, the expression \eqref{osbssi} for $B_{\mu\nu}$ therefore becomes:
\be
\begin{split}
   2B_{\mu\nu}=\Theta_{\mu\nu}\nabla_\alpha\phi\nabla^\alpha\phi+\Sigma_{\mu\nu}^{\sigma\theta}\nabla_\sigma\phi\nabla_\theta\phi\ .
\end{split}
\ee
Making the last terms explicit, we have:
\be
\begin{split}
    2B_{\mu\nu}&=\frac{1}{2}\Big[\delta_\nu^{\ \theta}\nabla_\mu\nabla^\sigma -g_{\mu\nu}\nabla^\sigma\nabla^\theta +\delta_{\mu}^{\ \sigma}\nabla^\theta\nabla_\nu\Big]\nabla_\sigma\phi\nabla_\theta\phi\\
&\quad-\frac{R}{4}\nabla_\mu\phi\nabla_\nu\phi+\frac{1}{2}\left(g_{\mu\nu}\Delta-\nabla_\mu\nabla_\nu \right)\nabla_\alpha\phi\nabla^\alpha\phi\\
&\quad-\Delta\left(\nabla_\mu\phi\nabla_\nu\phi\right)\ .
\end{split}
\ee
For the sake of simplicity, we can further assume the scalar field to be massless. Concerning $E_{\mu\nu}$, we are hence left with:
\be
\begin{split}
E_{\mu\nu}&=-\nabla_\mu\phi\nabla_\nu R-\nabla_\nu\phi\nabla_\mu R-\frac{3}{4}\nabla_\mu\phi\nabla_\nu\left(\nabla\phi\right)^{2}\\
    &\quad-\frac{3}{4}\nabla_\nu\phi\nabla_\mu\left(\nabla\phi\right)^{2}+\left(R_{\mu\nu}-\frac{1}{4}\nabla_\mu\phi\nabla_\nu\phi\right)\nabla_\alpha\phi\nabla^\alpha\phi\\
    &\quad+g_{\mu\nu}\left[\nabla^\alpha\phi\nabla_\alpha R+\frac{3}{4}\nabla^\alpha\phi\nabla_\alpha\left(\nabla\phi\right)^{2}\right]\ .
\end{split}
\ee
From such expressions, we can obtain $S_{\mu\nu}$ as $E_{\mu\nu}-B_{\mu\nu}$:
\be
\begin{split}\label{2ds}
    S_{\mu\nu}&=-\nabla_\mu\phi\nabla_\nu R-\nabla_\nu\phi\nabla_\mu R-\frac{3}{2}\nabla_\mu\phi\nabla^\alpha\phi\nabla_\nu\nabla_\alpha\phi\\
    &\quad-\frac{3}{2}\nabla_\nu\phi\nabla^\alpha\phi\nabla_\mu\nabla_\alpha\phi-\frac{1}{4}\nabla_\mu\phi\nabla_\nu\phi\nabla_\alpha\phi\nabla^\alpha\phi\\
    &\quad+g_{\mu\nu}\left[\nabla^\alpha\phi\nabla_\alpha R+\frac{3}{2}\nabla^\alpha\phi\nabla^\beta\phi\nabla_\alpha\nabla_\beta\phi\right]\\
&\quad+\frac{R}{4}\nabla_\mu\phi\nabla_\nu\phi-g_{\mu\nu}\left(\nabla^\beta\nabla_\beta\nabla_\alpha\phi\nabla^\alpha\phi+\nabla^\beta\nabla_\alpha\phi\nabla_\beta\nabla^\alpha\phi\right)\\
&\quad+\nabla^\alpha\nabla_\alpha\nabla_\mu\phi\nabla_\nu\phi+2\nabla^\alpha\nabla_\mu\phi\nabla_\alpha\nabla_\nu\phi\\
&\quad +\nabla_\mu\nabla_\nu\nabla_\alpha\phi\nabla^\alpha\phi+\nabla_\nu\nabla_\alpha\phi\nabla_\mu\nabla^\alpha\phi+\nabla_\mu\phi\nabla^\alpha\nabla_\alpha\nabla_\nu\phi\\
&\quad-\frac{1}{2}\nabla_\mu\phi\nabla^\theta\nabla_\nu\nabla_\theta\phi-\frac{1}{2}\nabla_\sigma\phi\nabla_\mu\nabla^\sigma\nabla_\nu\phi\\
&\quad -\frac{1}{2}\nabla^\theta\nabla_\nu\nabla_\mu\phi\nabla_\theta\phi-\frac{1}{2}\nabla_\nu\nabla_\mu\phi\nabla^\theta\nabla_\theta\phi-\frac{1}{2}\nabla^\theta\nabla_\mu\phi\nabla_\nu\nabla_\theta\phi\\
&\quad-\frac{1}{2}\nabla_\mu\nabla^\sigma\nabla_\sigma\phi\nabla_\nu\phi-\frac{1}{2}\nabla^\sigma\nabla_\sigma\phi\nabla_\mu\nabla_\nu\phi-\frac{1}{2}\nabla_\mu\nabla_\sigma\phi\nabla^\sigma\nabla_\nu\phi\\
&\quad+\frac{1}{2}g_{\mu\nu} \Big(\nabla^\sigma\nabla^\theta\nabla_\sigma\phi\nabla_\theta\phi+\nabla^\theta\nabla_\sigma\phi\nabla^\sigma\nabla_\theta\phi+\nabla^\sigma\nabla_\sigma\phi\nabla^\theta\nabla_\theta\phi\\
&\quad+\nabla_\sigma\phi\nabla^\sigma\nabla^\theta\nabla_\theta\phi\Big)\ .
\end{split}
\ee
Clearly, the expression \eqref{2ds} is still too complicated to be treated nicely. In order to simplify it, we refer to the explicit computations performed in \ref{app2d}. Such derivations allow us to observe that $\nabla_\alpha\nabla_\beta\phi=\nabla_\beta\nabla_\alpha\phi$. By exploiting such symmetry, we get:
\be
\begin{split}\label{2dssimp}
    S_{\mu\nu}&=-\nabla_\mu\phi\nabla_\nu R-\nabla_\nu\phi\nabla_\mu R-\frac{3}{2}\nabla_\mu\phi\nabla^\alpha\phi\nabla_\nu\nabla_\alpha\phi\\
    &\quad-\frac{3}{2}\nabla_\nu\phi\nabla^\alpha\phi\nabla_\mu\nabla_\alpha\phi-\frac{1}{4}\nabla_\mu\phi\nabla_\nu\phi\nabla_\alpha\phi\nabla^\alpha\phi\\
    &\quad+g_{\mu\nu}\left(\nabla^\alpha\phi\nabla_\alpha R+\frac{3}{2}\nabla^\alpha\phi\nabla^\beta\phi\nabla_\alpha\nabla_\beta\phi\right)\\
&\quad+\frac{R}{4}\nabla_\mu\phi\nabla_\nu\phi+\Delta\nabla_\mu\phi\nabla_\nu\phi+2\nabla^\alpha\nabla_\mu\phi\nabla_\alpha\nabla_\nu\phi\\
&\quad +\frac{1}{2}\nabla_\mu\nabla_\nu\nabla_\alpha\phi\nabla^\alpha\phi+\frac{1}{2}\nabla_\mu\phi\Delta\nabla_\nu\phi\\
&\quad -\frac{1}{2}\nabla^\theta\nabla_\nu\nabla_\mu\phi\nabla_\theta\phi-\nabla_\nu\nabla_\mu\phi\Delta\phi-\frac{1}{2}\nabla_\mu\Delta\phi\nabla_\nu\phi\\
&\quad+\frac{1}{2}g_{\mu\nu} \Big(\Delta\phi\Delta\phi+\nabla_\sigma\phi\nabla^\sigma\Delta\phi-\Delta\nabla^\theta\phi\nabla_\theta\phi-\nabla^\theta\nabla_\sigma\phi\nabla^\sigma\nabla_\theta\phi\Big)\ .
\end{split}
\ee
By assuming both $R$ and the initial scalar configuration to be constant, we have that such conditions are conserved along the whole flow. So, we get that $S_{\mu\nu}$ simplifies to:
\be
\begin{split}
    S_{\mu\nu}&=0\ .
\end{split}
\ee
At this point, we an explicitly solve the flow equations. We do so by taking the metric to be that of $2$-dimensional (Anti-)de Sitter spacetime
\be 
g_{1}=1-\frac{\kappa x^2}{2}\ ,\quad g_{2}=\frac{1}{g_1}
\ee 
for which we simply have:
\be 
R=\kappa\ .
\ee
From the flow equation \eqref{specflmet} and the assumption of having a spacetime-constant scalar field, we can derive the flow equation
\be 
\frac{\de R}{\de\lambda}=R^2
\ee
for the Ricci scalar $R$, which can be solved explicitly as:
\be 
R\left(\lambda\right)=\frac{R_0}{1-R_0\lambda}\ .
\ee
As is well-known from previous works \cite{Kehagias:2019akr,DeBiasio:2020xkv}, it is well known that:
\begin{itemize}
    \item When $R_0>0$, the evolution reaches a singularity at a finite flow time.
    \item When $R_0<0$, the evolution asymptotically flows to flat spacetime.
\end{itemize}
Thus, we have:
\be 
\kappa\left(\lambda\right)=\frac{\kappa_0}{1-\kappa_0\lambda}\ .
\ee 
Concerning the scalar flow equation
\be 
\frac{\de\phi}{\de\lambda}=-R\ ,
\ee 
we can solve it and get:
\be 
\phi\left(\lambda\right)=\phi_0+\int_0^\lambda\frac{\kappa_0}{\kappa_0\tau-1}\diff\tau=\phi_0+\log{\left(1-\kappa_0\lambda\right)}\ .
\ee 
Since $S_{\mu\nu}=0$, the $2$-dimensional example gives us no interesting behaviour. This is something that should have been expected. Indeed, general relativity is trivial in $D=2$ and the energy-momentum tensor for a constant scalar is null.

\subsubsection{An example in $D=3$}\label{specific2}
\begin{figure}[H]
    \centering
    \includegraphics[width=0.8\linewidth]{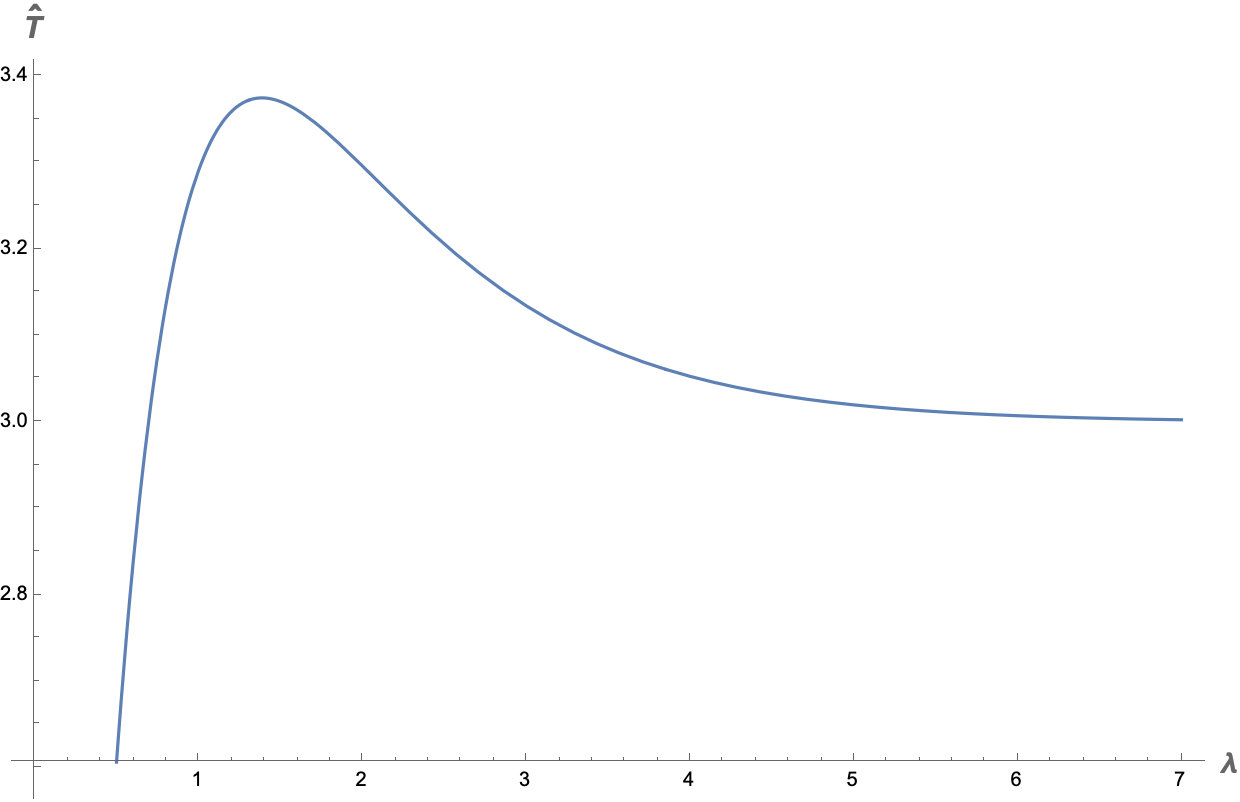}
    \caption{Plot of $\hat{T}$ against the flow parameter $\lambda$.}
    \label{fig}
\end{figure}

In the following discussion, we reproduce the analysis performed in \ref{specific} in the case in which $D=3$. In order to do so, we once more assume to work with an initial condition comprised of the metric of an Einstein manifold -more specifically, that of Anti de-Sitter spacetime in $D=3$- and a massless, scalar field with constant spacetime value $\phi_0$. 
By assuming, once more, $d=b=0$ in the flow equations, we are left with:
\be
\begin{split}
 \frac{\de g_{\mu\nu}}{\de\lambda}= -2R_{\mu\nu}+\frac{1}{2}\nabla_\mu\phi\nabla_\nu\phi\ ,\quad   
  \frac{\de\phi}{\de\lambda}=-R-\frac{3}{4}\left(\nabla\phi\right)^{2}\ .
\end{split}
\ee
Such equations conserve our assumptions on $g_{\mu\nu}$ and $\phi$, so that we get
\be
\frac{\de R}{\de\lambda}=\frac{2R^2}{3}\ ,\quad \frac{\de\phi}{\de\lambda}=-R\ ,
\ee
precisely as in \ref{specific}, except for a small deviation due to different dimensionality. 
By already removing the terms containing derivatives of $\phi$, plugging-in the expression of $R_{\mu\nu}$ in terms of $R$ and $g_{\mu\nu}$ and remembering Anti-de Sitter spacetime scalar curvature is linked to the cosmological constant $\Lambda$ by
\be \label{equal}
R=6\Lambda
\ee
the expression for $S_{\mu\nu}$ gets to be:
\be
\begin{split}
   S_{\mu\nu}&=\left(4\Lambda^2-C\right)g_{\mu\nu}\ .
\end{split}
\ee
Therefore, the flow equation for the emergent energy-momentum tensor $\hat{T}_{\mu\nu}$ is:
\be
\frac{\de\hat{T}_{\mu\nu}}{\de\lambda}=\left(4\Lambda^2-C\right)g_{\mu\nu}\ .
\ee
Plugging-in an ansatz of the form
\be
\hat{T}_{\mu\nu}=\frac{\hat{T}}{3}g_{\mu\nu}\ ,
\ee
our equation simply reduces to:
\be
\frac{\de\hat{T}}{\de\lambda}=3\left(4\Lambda^2-C\right)\ .
\ee
By imposing the initial condition $\hat{T}_{\mu\nu}=0$ and remembering that $C$ appeared as the flow-source for $\Lambda$, we can solve the above equation as:
\be
\hat{T}\left(\lambda\right)=12\int_0^\lambda\Lambda^2\left(\tau\right)\diff\tau-3\left[\Lambda\left(\lambda\right)-\Lambda_0\right]\ .
\ee
In the following, we will explore three natural behaviours for $\Lambda$. First of all, we can take it to be constant in the flow parameter. Hence, we have:
\be
\hat{T}\left(\lambda\right)=12\Lambda^2\lambda\ .
\ee
If we instead take $\Lambda$ to be exponentially dropping as
\be 
\Lambda\left(\lambda\right)=\Lambda_0e^{-\lambda}\ ,
\ee
we would get:
\be
\hat{T}\left(\lambda\right)=6\Lambda_0^2\left(1-e^{-2\lambda}\right)+3\Lambda_0\left(1-e^{-\lambda}\right)\ .
\ee
For large values of the flow parameter, we get:
\be 
\hat{T}\left(\lambda\right)=3\Lambda_0\left(2\Lambda_0+1\right)\ .
\ee 
By assuming $\Lambda_0=-1$, this would imply the behaviour depicted in figure \ref{fig}. 
Last, we must consider the case in which the \eqref{equal} is taken to hold for the whole flow. Thus, $S_{\mu\nu}=0$ and no extra energy-momentum tensor $\hat{T}_{\mu\nu}$ emerges. Generally speaking and assuming any emergent field to be a real scalar, an energy-momentum tensor satisfying $\hat{T}_{\mu\nu}=\frac{\hat{T}}{3}g_{\mu\nu}$ can only be achieved by imposing
\be
\hat{T}=-\sum_{k=1}^{M}\frac{3m_k^2}{2}\phi_k^2\ ,
\ee
where $\phi_k$ are $M$ emergent spacetime-constant fields, which we will from now on normalise to $\phi_k=\sqrt{2/3}$, and $m_k$ the corresponding mass scales. Thus, we get:
\be
\hat{T}=-\sum_{k=1}^{M}m_k^2\ .
\ee
Therefore, making such masses flow-dependent, we have:
\be 
\sum_{k=1}^{M}m_k^2\left(\lambda\right)=12\int_0^\lambda\Lambda^2\left(\tau\right)\diff\tau-3\left[\Lambda\left(\lambda\right)-\Lambda_0\right]\ .
\ee
Taking the masses to be distributed as 
\be
m_k=m\sqrt{n}
\ee
and assuming only a finite portion of them to enter the theory, we would get:
\be 
m^2\left(\lambda\right)=\frac{12}{C_M}\int_0^\lambda\Lambda^2\left(\tau\right)\diff\tau-\frac{3}{C_M}\left[\Lambda\left(\lambda\right)-\Lambda_0\right]\ ,
\ee
where the numerical constant is
\be
C_M\equiv \frac{M(M+1)}{2}
\ee
and depends on the number of extra fields we choose to include in the theory.

\subsection{Scalar field with sextic potential}\label{potentialcase}
\begin{figure}[H]
    \centering
    \includegraphics[width=0.8\linewidth]{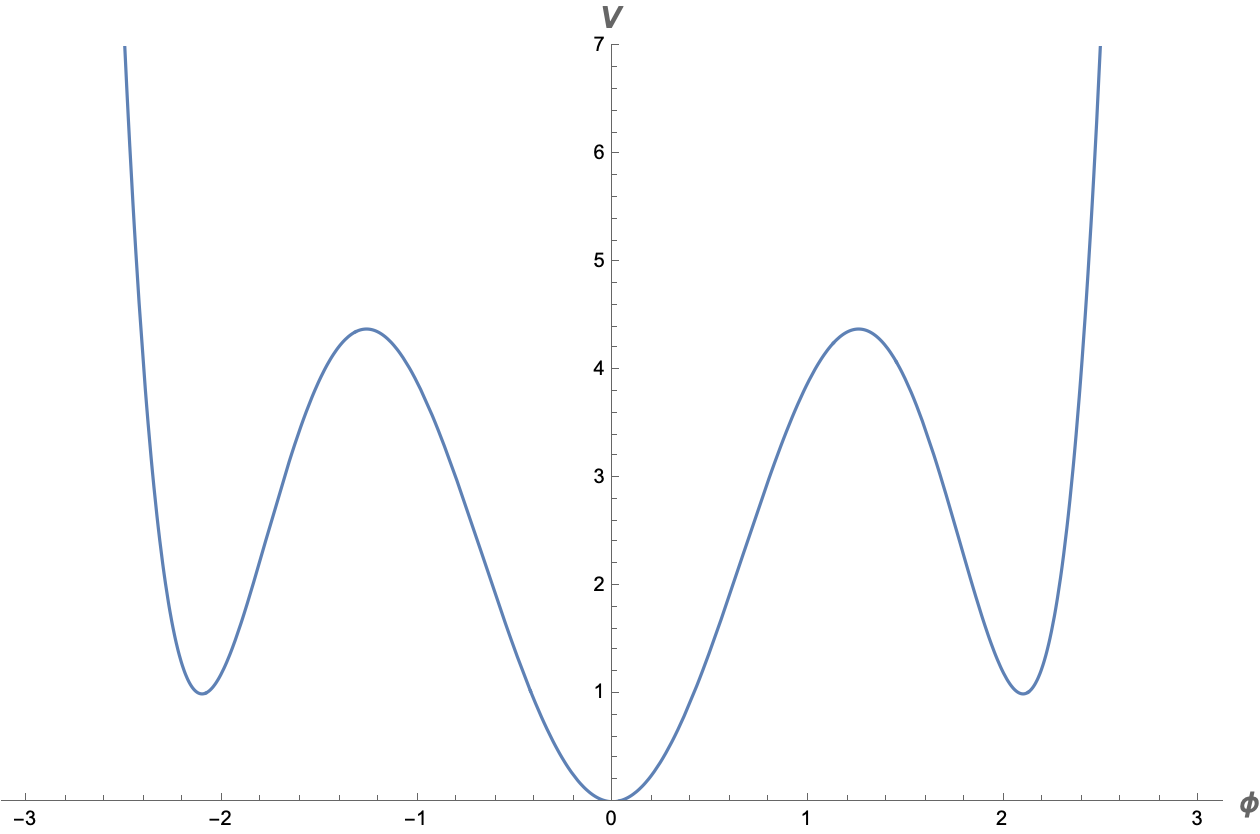}
    \caption{Plot of the behaviour of $V\left(\phi\right)$ around $\phi=0$, when $V_0=1$.}
    \label{pottV}
\end{figure}

\begin{figure}[H]
    \centering
    \includegraphics[width=0.8\linewidth]{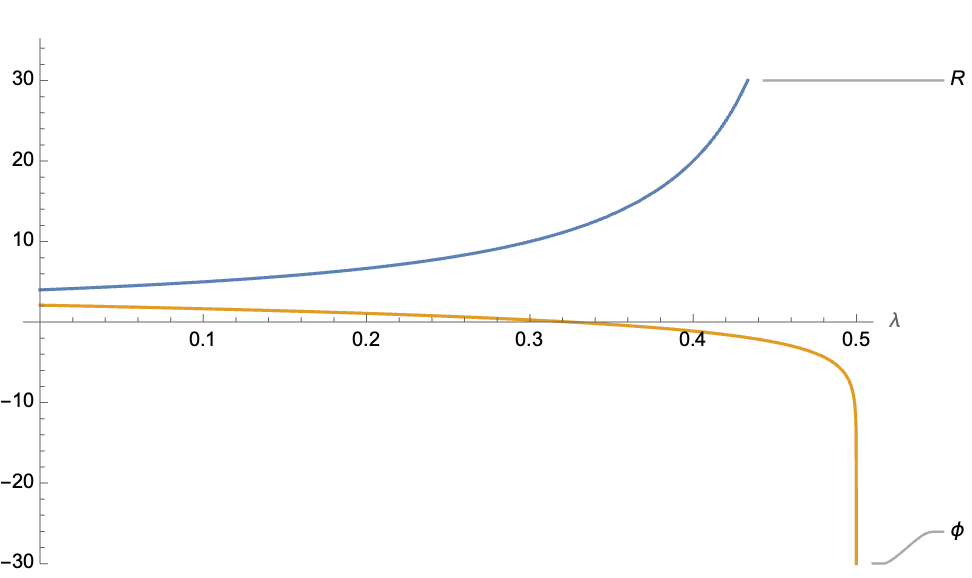}
    \caption{Plot of the flow behaviour of $R$ and $\phi$, in $D=4$. For $\lambda\to 0.5$, both $R$ and $\phi$ diverge (towards positive and negative values, respectively).}
    \label{flowbeh}
\end{figure}

\begin{figure}[H]
    \centering
    \includegraphics[width=0.9\linewidth]{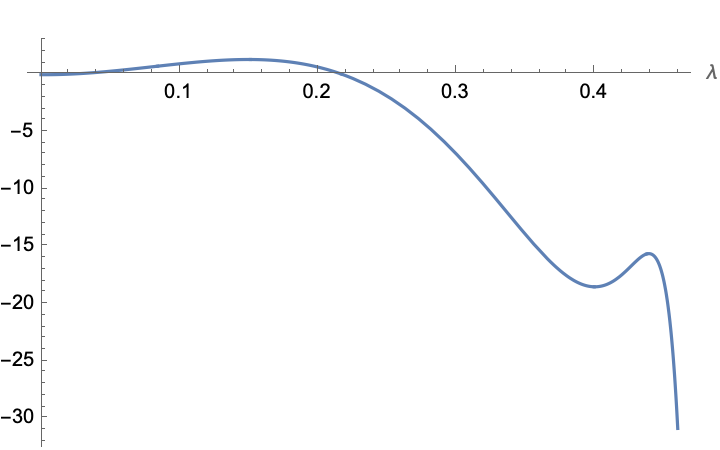}
    \caption{Plot of the flow behaviour of $\Omega$ for $D=4$ and $\lambda\in[0,0.46]$.}
    \label{czero}
\end{figure}

\begin{figure}[H]
    \centering
    \includegraphics[width=0.9\linewidth]{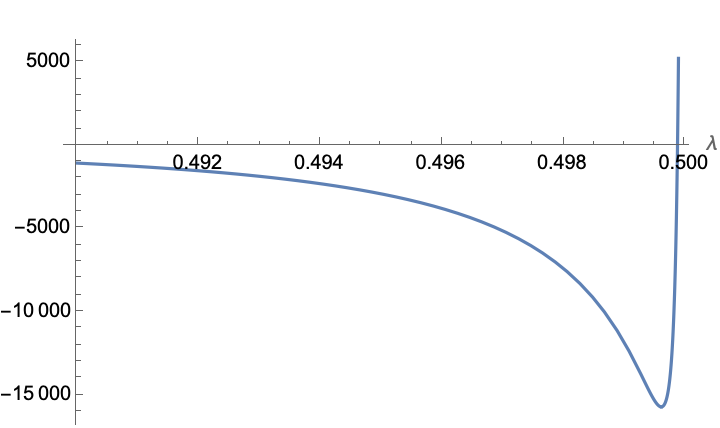}
    \caption{Plot of the flow behaviour of $\Omega$ for $D=4$ and $\lambda\in[0.49,0.4999]$. For $\lambda\to 0.5$, $\Omega$ blows to $+\infty$.}
    \label{czero0}
\end{figure}

In the following discussion, we assume to work with a single scalar field $\phi$, subject to a potential of the form
\be \label{6pot}
V\left(\phi\right)=\frac{V_0}{\alpha_0}\phi^2\left(\phi^4-9\phi^2+21\right)\ ,
\ee
with $\alpha_0=9-4\sqrt{2}$, so that the values of the potential minima of our interest are normalised to $V_0$, which is taken to be a positive parameter. The potential is shown in figure \ref{pottV}.
This can be achieved as a simple and straightforward generalisation of the discussion developed at the beginning of section \ref{scaler}, in which the only thing to be altered is the specific form of the energy-momentum tensor:
\be
T_{\mu\nu}=\nabla_\mu\phi\nabla_\nu\phi-g_{\mu\nu}\left[\frac{1}{2}\nabla^\alpha\phi\nabla_\alpha\phi+V\left(\phi\right)\right]\ .
\ee
The equations of motion for $\phi$ can be solved by taking the scalar to have the spacetime-constant value $\phi_0$, corresponding to a minimum of the potential. This will serve as our initial flow configuration, together with an appropriate solution for the metric. Since the potential will produce an effective cosmological constant term
\be
\Lambda_{\text{eff}}=V_0\ ,
\ee
we take $\Lambda=C=0$ and simplify the whole discussion. Therefore, our initial metric will be that of $D$-dimensional de Sitter spacetime with scalar curvature:
\be 
R=\frac{2D}{D-2}V_0\ .
\ee
Since we have taken the scalar to be spacetime constant at the starting point of the flow, both this condition and the metric being that of de Sitter spacetime are conserved along the flow. Hence, by taking $d=b=0$, the flow equations reduce to:
\be 
\frac{\de R}{\de\lambda}=\frac{2R^2}{D}\ ,\quad \frac{\de\phi}{\de\lambda}=-R\ .
\ee 
By solving the equations, as we did in the previous examples, we get
\be 
R\left(\lambda\right)=\frac{DR_0}{D-R_02\lambda}\ ,\quad \phi\left(\lambda\right)=\phi_0+\frac{D}{2}\log\left(1-\frac{2R_0\lambda}{D}\right)\ ,
\ee
where we have defined
\be 
\phi_0\equiv\sqrt{3+\sqrt{2}}\ ,
\ee
so that it corresponds to the potential minimum $V\left(\phi_0\right)=V_0$. For the rest of the discussion, we take $V_0=1$ and have our flow behaviours reduce to:
\be 
R\left(\lambda\right)=\frac{2D^2}{D(D-2)-4D\lambda}\ ,\quad \phi\left(\lambda\right)=\phi_0+\frac{D}{2}\log\left(1-\frac{4\lambda}{D-2}\right)\ ,
\ee
Figure \ref{flowbeh} shows them explicitly against $\lambda$ in $D=4$.
It can be straightforwardly shown that:
\be
\begin{split}
   E_{\mu\nu}&=-g_{\mu\nu}\frac{\de V\left(\varphi\right)}{\de\varphi}\Bigg|_{\varphi=\phi}\cdot\frac{\de\phi}{\de\lambda}-\frac{\de g_{\mu\nu}}{\de\lambda}V\left(\phi\right) \\
   &=\frac{2R\phi}{D\alpha_0}g_{\mu\nu}\left(\phi^5+3D\phi^4-9\phi^3-18D\phi^2+21\phi+21D\right)\ .
\end{split}
\ee
\be 
\begin{split}
    B_{\mu\nu}&=0\ .
\end{split}
\ee
Concerning the flow of the extra energy-momentum tensor term, we hence obtain
\be
S_{\mu\nu}=C_0g_{\mu\nu}\ ,
\ee
where we have defined:
\be
\begin{split}
    C_0&=-\frac{2R\phi}{D\alpha_0}\left(\phi^5+3D\phi^4-9\phi^3-18D\phi^2+21\phi+21D\right)\ .
\end{split}
\ee
\begin{figure}[H]
    \centering
    \includegraphics[width=0.9\linewidth]{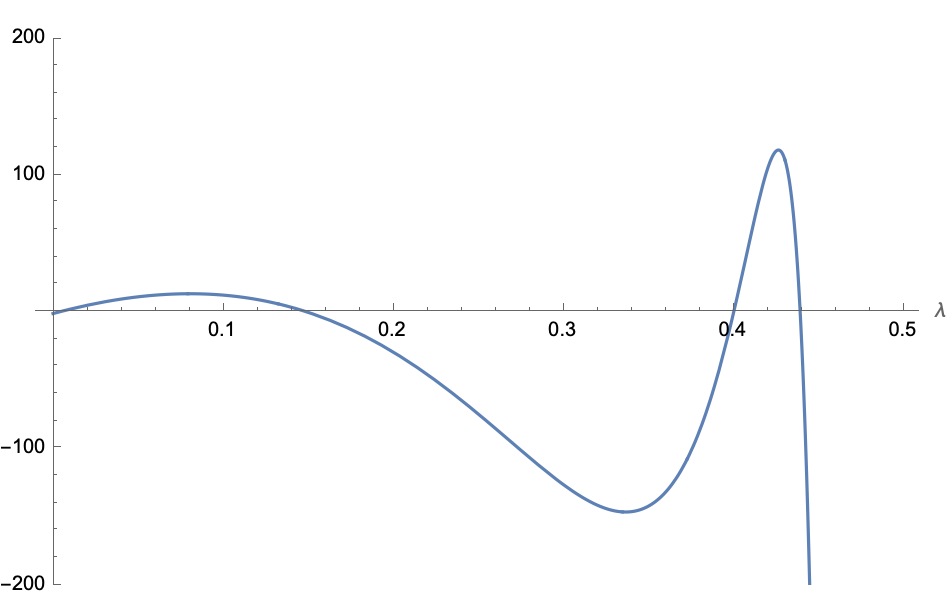}
    \caption{Plot of the flow behaviour of $C_0$ for $D=4$.}
    \label{czer}
\end{figure}
At this point, if we can take the extra energy momentum tensor to be of the extremely simple diagonal form
\be
\hat{T}_{\mu\nu}=\Omega g_{\mu\nu}\ ,
\ee
$\Omega$ follows the flow equation:
\be 
\frac{\de\Omega}{\de\lambda}=C_0\ .
\ee
The flow equation can be solved numerically, obtaining the plot shown in figures \eqref{czero} and \eqref{czero0}. For $\lambda\to 0.5$, $\Omega$ blows to $+\infty$.

\subsubsection{Flow equations from the action}\label{fromact}
\begin{figure}[H]
    \centering
    \includegraphics[width=\linewidth]{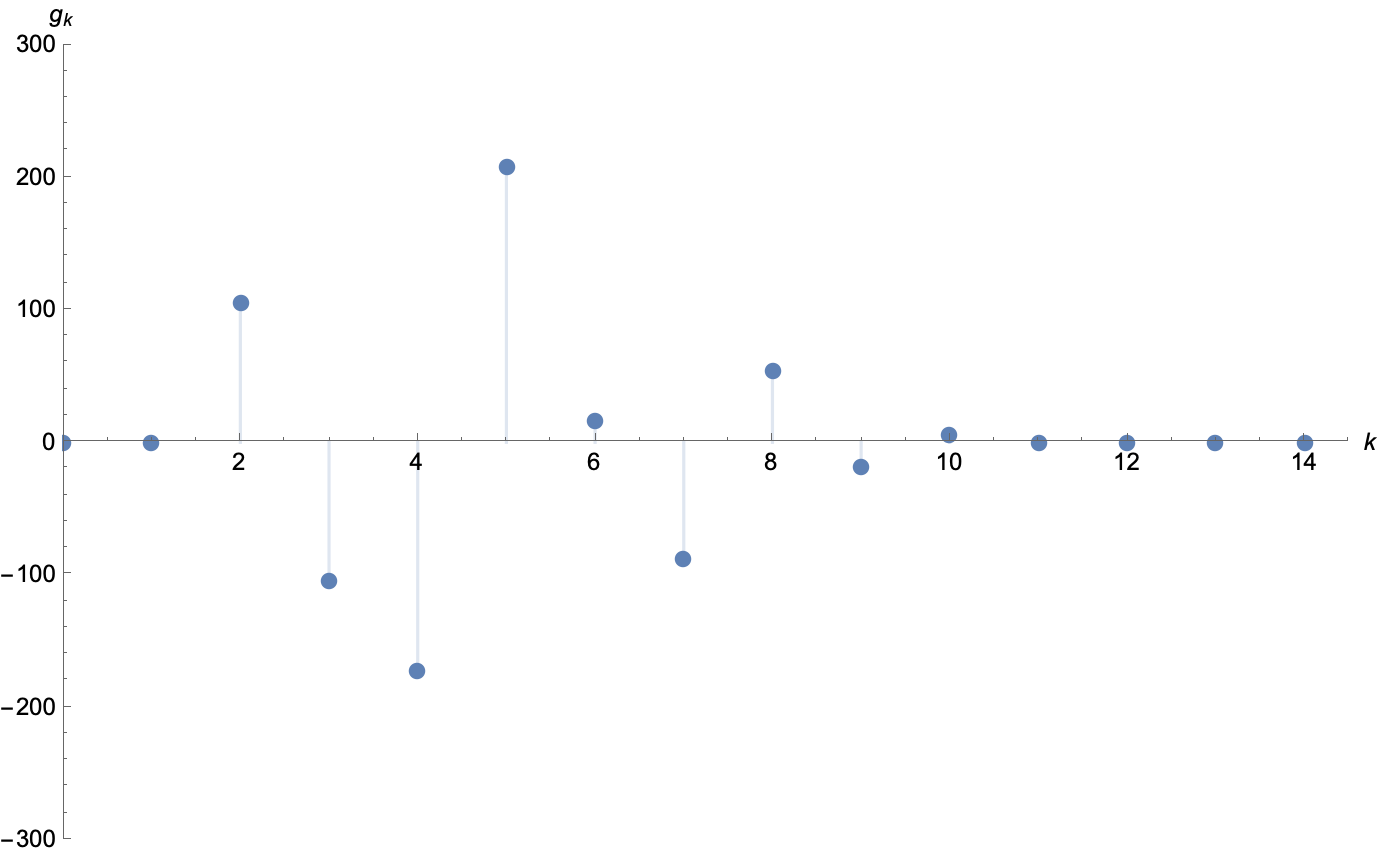}
    \caption{Plot of $g_{\ k}^{(4)}$ for different values of $k$, at $D=4$. Clearly, the coefficients drop to zero in an extremely quick way.}
    \label{ddott}
\end{figure}

\begin{figure}[H]
    \centering
    \includegraphics[width=0.9\linewidth]{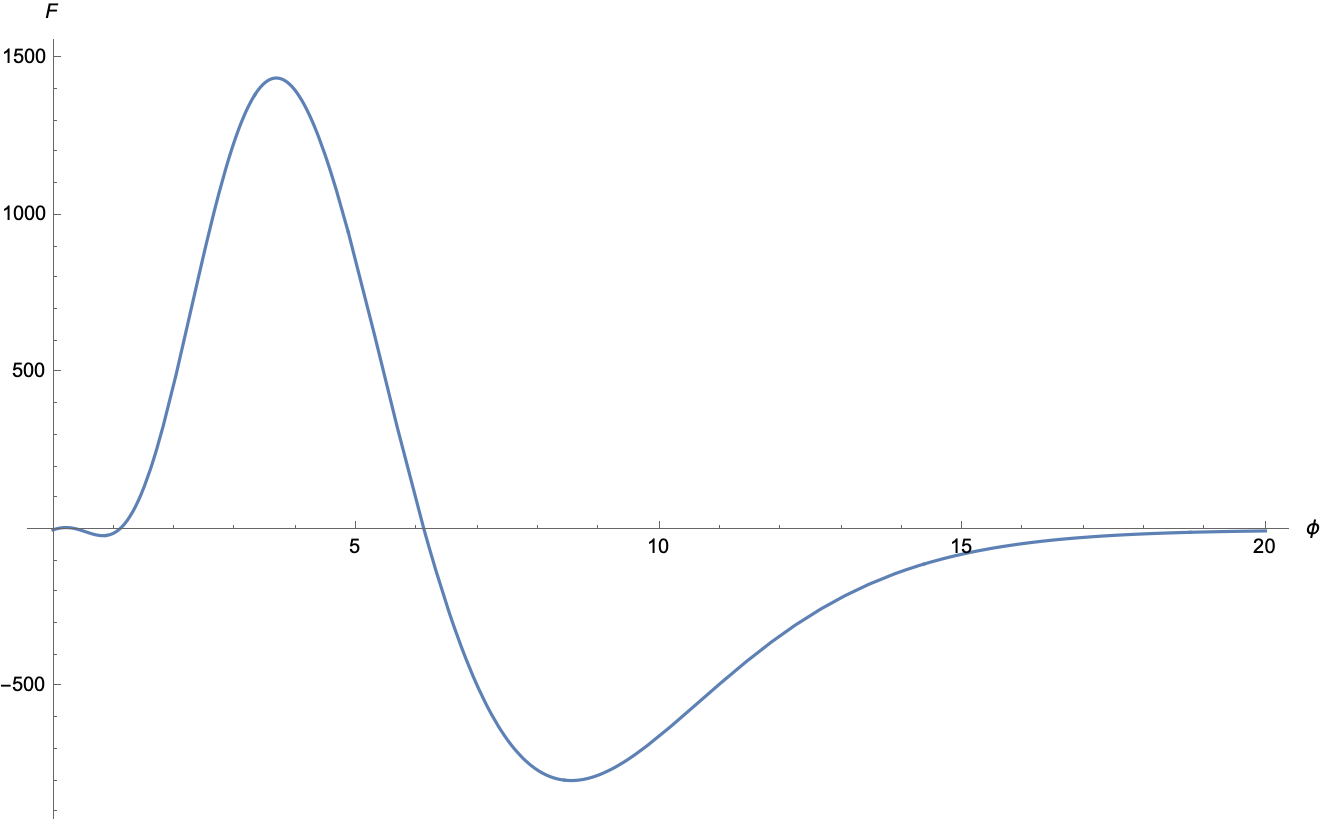}
    \caption{Plot of F against $\phi$, in which the exponential suppression can be observed.}
    \label{Fpl}
\end{figure}

\begin{figure}[H]
    \centering
    \includegraphics[width=0.9\linewidth]{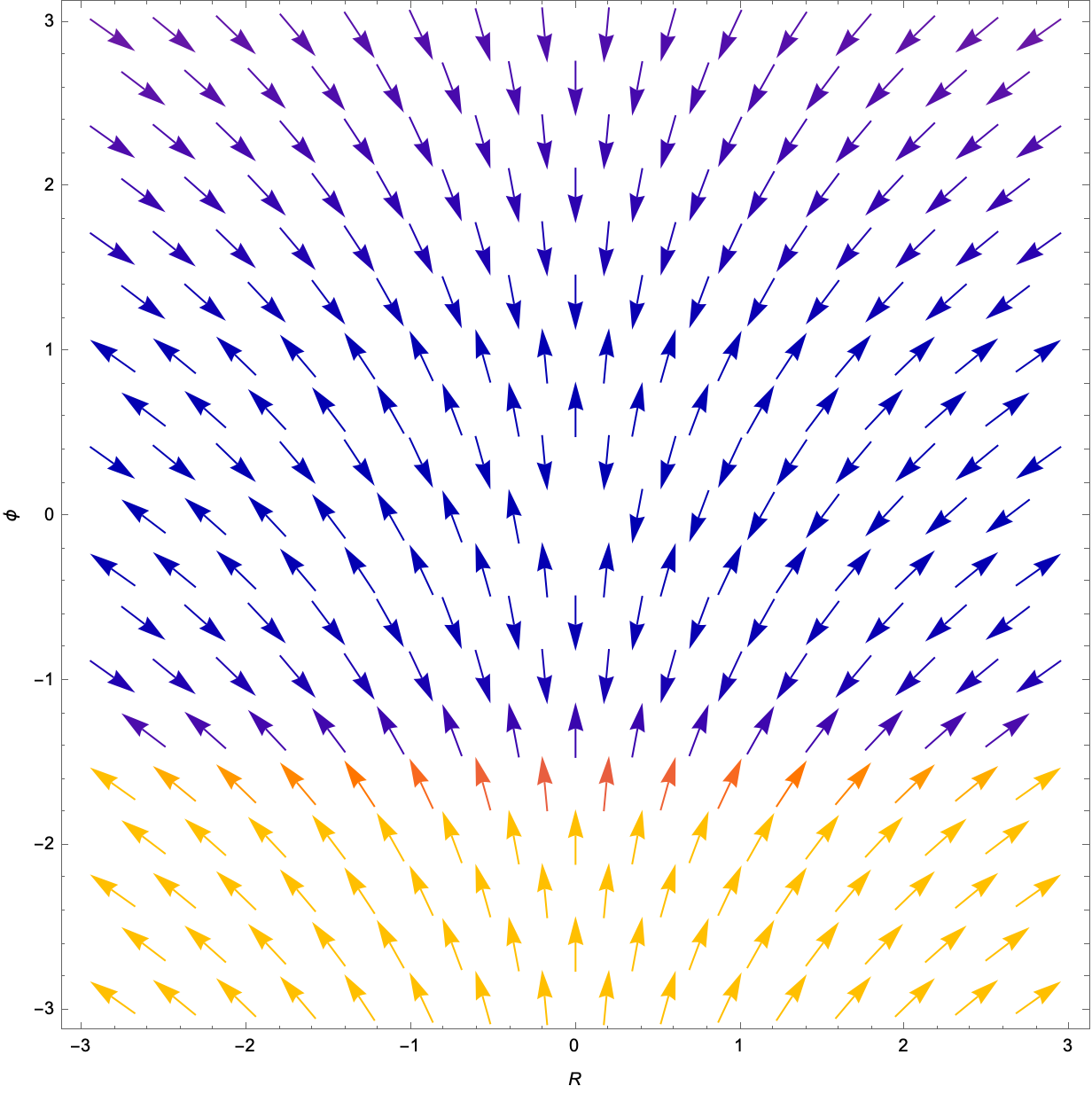}
    \caption{Plot of the vector field tangent to the flow, in the $\left(R,\phi\right)$ space.}
    \label{vec}
\end{figure}

\begin{figure}[H]
    \centering
    \includegraphics[width=0.9\linewidth]{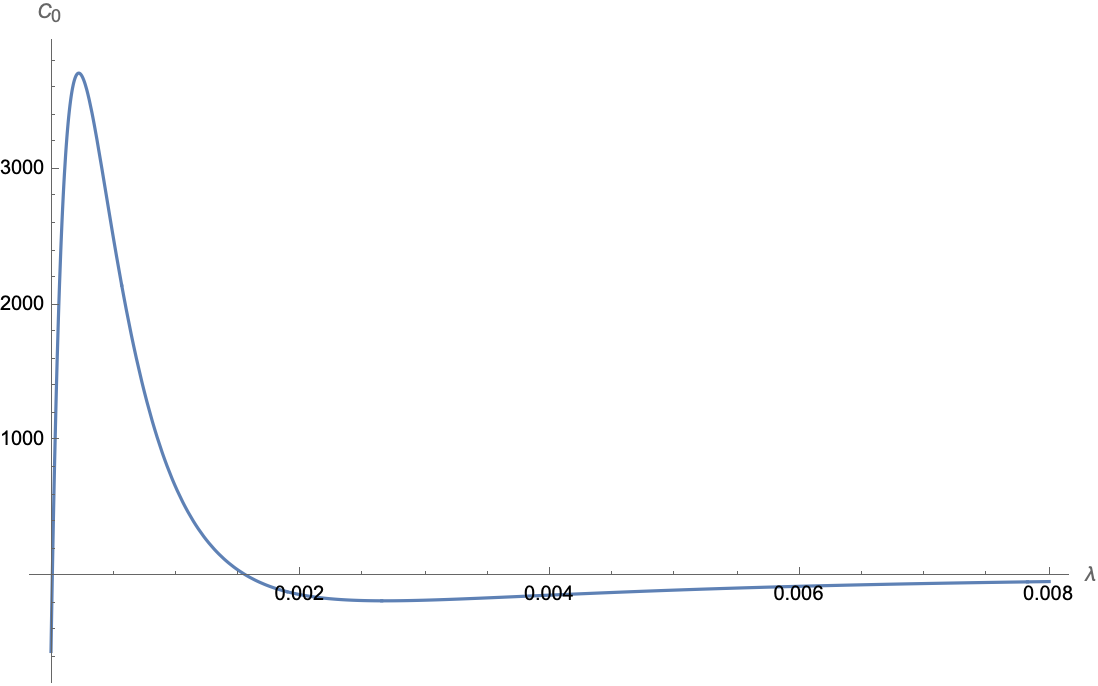}
    \caption{Plot of $C_0$ against the flow parameter $\lambda$.}
    \label{ssoss}
\end{figure}

In the following discussion, we will once more consider the initial metric and scalar configuration introduced in \ref{potentialcase}. Namely, we work with a $D$-dimensional Anti-de Sitter spacetime geometry, whose curvature is sourced by the spacetime constant value of a scalar $\phi$, sitting at the appropriate minimum of the potential \eqref{6pot}. Indeed, we have
\be 
\phi_0\equiv\sqrt{3+\sqrt{2}}\ ,\quad R_0=\frac{2D}{D-2}\ ,
\ee
where we have once more taken $V_0=1$. Moreover, we assume $D>2$. The main change with respect to the previous analysis is that we consider a different set of flow equations. In particular, exploiting the general results described in \cite{DeBiasio:2022omq}, we derive a system of gradient flow equations from a new entropy-$\mathcal{F}$ functional, obtained as the euclidean string-frame version of the action of our theory. This way, the introduction of the potential \eqref{6pot} directly influences the expression for the flow equations. In the above-mentioned reference, it was shown that an action
\begin{equation}\label{aact}
    S=\int\diff^{D}x\sqrt{\tilde{g}}\left[\tilde{R}+\frac{1}{2}\left(\nabla \phi\right)^{2}-\sum_{n=0}^{+\infty}\frac{g_{n}}{n!}\phi^{n}\right]\ .
\end{equation}
naturally induces, neglecting the diffeomorphism term, the flow equations
\begin{equation}\label{acteq}
    \begin{split}
        \frac{\de g_{\mu\nu}}{\de\lambda}=&-2R_{\mu\nu}-g_{\mu\nu}\sum_{n=1}^{+\infty}ns_{\ n}^{(D)}\phi^{n-1}+4\frac{D-1}{D-2}\nabla_{\mu}\phi\nabla_{\nu}\phi\\
        &-2\frac{5D-6}{D-2}\nabla_{\nu}\nabla_{\mu}\phi-2\frac{D-1}{D-2}g_{\mu\nu}\nabla^{2}\phi+2\frac{D-1}{D-2}g_{\mu\nu}\left(\nabla \phi\right)^2\ ,\\
        \frac{\de \phi}{\de\lambda}=&-R-\frac{D}{2}\sum_{n=1}^{+\infty}ns_{\ n}^{(D)} \phi^{n-1}+\frac{\left(D-1\right)\left(D+2\right)}{D-2}\left(\nabla \phi\right)^{2}\\
        &-\frac{D^{2}+4D-6}{D-2}\nabla^{2} \phi\ ,
    \end{split}
\end{equation}
where the $s_{\ k}^{(D)}$ parameters have been defined as:
\be 
    s_{\ k}^{(D)}\equiv\sum_{n=0}^{k}\frac{g_{n}}{\left(k-n\right)!n!}\left(\frac{2}{2-D}\right)^{k-n}\left(\frac{4D-6}{D-2}\right)^{n/2}\ .
\ee
For the specific case of the potential
\be
V\left(\phi\right)=\frac{1}{\alpha_0}\phi^2\left(\phi^4-9\phi^2+21\right)\ ,
\ee
with $\alpha_0=9-4\sqrt{2}$, we have:
\be
g_2=\frac{42}{\alpha_0}\ ,\quad g_4=-\frac{9\cdot4!}{\alpha_0}\ ,\quad g_6=\frac{6!}{\alpha_0}\ .
\ee
Hence, the $s_{\ k}^{(D)}$ parameters are:
\be
\begin{split}
    s_{\ 0}^{(D)}&=s_{\ 1}^{(D)}=0\ ,\quad s_{\ 2}^{(D)}=\frac{42}{\alpha_0}\cdot\frac{2D-3}{D-2}\ ,\quad
    s_{\ 3}^{(D)}=-\frac{84}{\alpha_0}\cdot\frac{2D-3}{\left(D-2\right)^2}\ ,\\ s_{\ 4}^{(D)}&=-\frac{12}{\alpha_0}\cdot\frac{2D-3}{\left(D-2\right)^3}\left(6D^2-21D+11\right)\ ,\\
    s_{\ 5}^{(D)}&=\frac{8}{\alpha_0}\cdot\frac{2D-3}{(D-2)^4}\left(18D^2-27D-36D+47\right)\ ,\\
    s_{\ k\geq 6}^{(D)}&=\frac{(-1)^k2^{k-3}}{\left(k-6\right)!\alpha_0}\cdot\frac{2D-3}{(D-2)^{k-3}}\Bigg[
    \frac{84\left(k-6\right)!}{\left(k-2\right)!(D-2)^2}\\
    &\quad-\frac{9\left(k-6\right)!\left(4D-6\right)}{\left(k-4\right)!(D-2)}+\left(2D-3\right)^{2}\Bigg]\ .
\end{split}
\ee
Defining the simpler parameters
\be 
g_{\ k}^{(D)}\equiv\alpha_0s_{\ k}^{(D)}
\ee
and taking $D=4$ for the sake of simplicity, we get:
\be
\begin{split}
    g_{\ 0}^{(4)}&=g_{\ 1}^{(4)}=0\ ,\quad g_{\ 2}^{(4)}=105\ ,\quad
    g_{\ 3}^{(4)}=-105\ ,\\ g_{\ 4}^{(4)}&=-\frac{345}{2}\ ,\quad g_{\ 5}^{(4)}=\frac{415}{2}\ ,\\
g_{\ k\geq6}^{(4)}&=(-1)^k\Bigg[
    \frac{105}{\left(k-2\right)!}-\frac{225}{\left(k-4\right)!}+\frac{125}{\left(k-6\right)!}\Bigg]\ .
\end{split}
\ee
The explicit $k$-behaviour of such factors is shown in figure \ref{ddott}.
At this point, we must compute the value of the function:
\be 
F\left(\phi\right)\equiv \sum_{n=1}^{+\infty}ns_{\ n}^{(4)} \phi^{n-1}\ .
\ee
Therefore, we have:
\be 
\begin{split}
    \alpha_0F\left(\phi\right)&=210\phi-315\phi^2-690\phi^3+\frac{2075}{2}\phi^4\\
    &\quad+\sum_{k=6}^{+\infty}k(-1)^k\Bigg[
    \frac{105}{\left(k-2\right)!}-\frac{225}{\left(k-4\right)!}+\frac{125}{\left(k-6\right)!}\Bigg]\phi^{k-1}\\
    &=210\phi-315\phi^2-690\phi^3+\frac{2075}{2}\phi^4\\
    &\quad+105\phi\left[2\sum_{n=4}^{+\infty}\frac{(-\phi)^n}{n!}-\phi\sum_{l=3}^{+\infty}
    \frac{(-\phi)^l}{l!}\right]\\
    &\quad-225\phi^3\left[4\sum_{n=2}^{+\infty}\frac{(-\phi)^n}{n!}-\phi\sum_{l=1}^{+\infty}\frac{(-\phi)^{l}}{l!}\right]\\
    &\quad+125\phi^5\left[6\sum_{n=0}^{+\infty}\frac{(-\phi)^n}{n!}-\phi\sum_{l=0}^{+\infty}\frac{(-\phi)^l}{l!}\right]\\
    &=210\phi-315\phi^2-690\phi^3+\frac{2075}{2}\phi^4\\
    &\quad+105\phi\left[\left(2-\phi\right)\left(e^{-\phi}-1+\phi-\frac{\phi^2}{2}\right)+\frac{\phi^3}{3}\right]\\
    &\quad-225\phi^3\left[\left(4-\phi\right)\left(e^{-\phi}-1\right)+4\phi\right]+125\phi^5\left(6-\phi \right)e^{-\phi}\\
   &=5\left[42\phi-21\phi^2-220\phi^3+55\phi^4+150\phi^5-25\phi^6\right]e^{-\phi}\ .
\end{split}
\ee
In the end, we get:
\be
F\left(\phi\right)=\frac{5}{\alpha_0}\left[42\phi-21\phi^2-220\phi^3+55\phi^4+150\phi^5-25\phi^6\right]e^{-\phi}\ .
\ee
The plot of such a function is given in figure \ref{Fpl}. Given that
we start with constant curvature and scalar field, it can be shown that the flow equations \eqref{acteq} preserve such conditions. Thus, they reduce to:
\begin{equation}
    \begin{split}
        \frac{\de R}{\de\lambda}=&-\frac{R}{2}\left[R+2F\left(\phi\right)\right]\ ,\\
        \frac{\de \phi}{\de\lambda}=&-\left[R+2F\left(\phi\right)\right]\ .
    \end{split}
\end{equation}
\begin{figure}[H]
    \centering
    \includegraphics[width=0.9\linewidth]{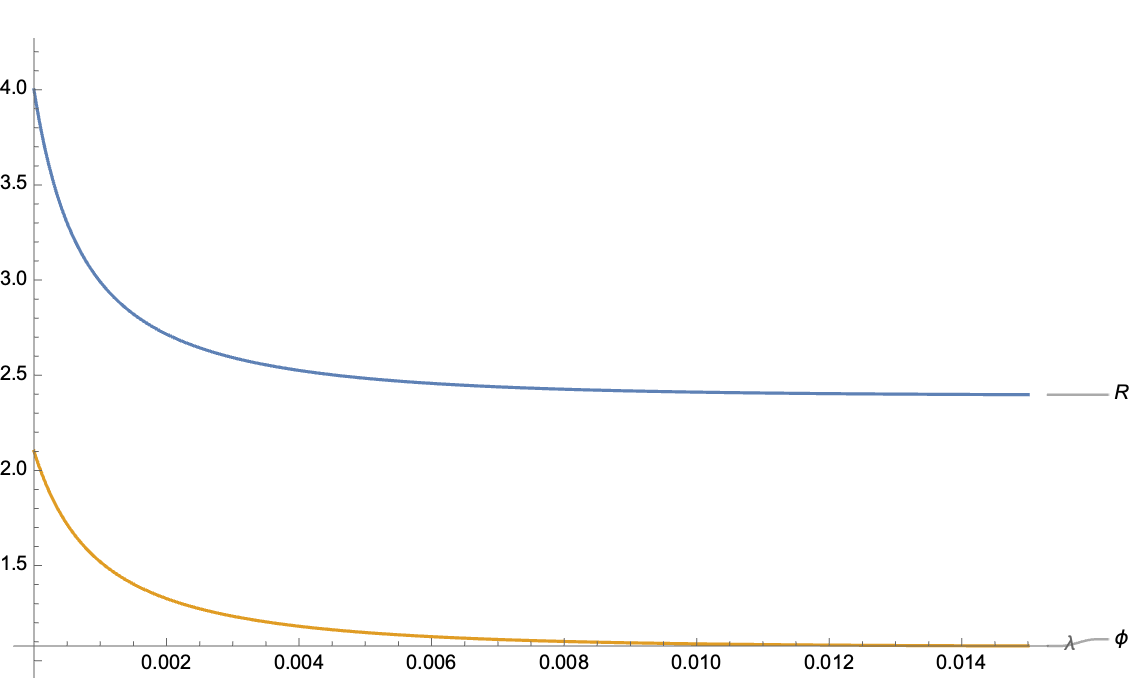}
    \caption{Flow behaviour for $R$ and $\phi$.}
    \label{fflo}
\end{figure}
Given the $D=4$ initial conditions
\be 
\phi_0\equiv\sqrt{3+\sqrt{2}}\ ,\quad R_0=4\ ,
\ee
the flow can be straightforwardly solved numerically, as shown in figure \ref{fflo}. More generally, figure \ref{vec} shows the vector fields tangent to the flow, in the $\left(R,\phi\right)$ space.
As in the previous derivation, we have $B_{\mu\nu}=0$ and
\be 
S_{\mu\nu}=C_0g_{\mu\nu}\ ,
\ee
with:
\be
\begin{split}
   C_0&=-\frac{1}{\alpha_0 }\left(\phi^6+12\phi^5-9\phi^4-72\phi^3+21\phi^2+84\phi\right)\left[\frac{R}{2}+F\left(\phi\right)\right]\ .
\end{split}
\ee
In figure \ref{ssoss}, the explicit flow behaviour of $C_0$ is shown. There we can observe that it goes to positive values, then to slightly negative ones and finally tends to zero. 
By introducing an extra energy-momentum tensor with the simple diagonal form 
\be 
\hat{T}_{\mu\nu}=\Omega g_{\mu\nu}\ ,
\ee
we have that $\Omega$ follows the simple flow equation:
\begin{equation}
    \frac{\de\Omega}{\de\lambda}=C_0\ .
\end{equation}
\begin{figure}[H]
    \centering
    \includegraphics[width=0.9\linewidth]{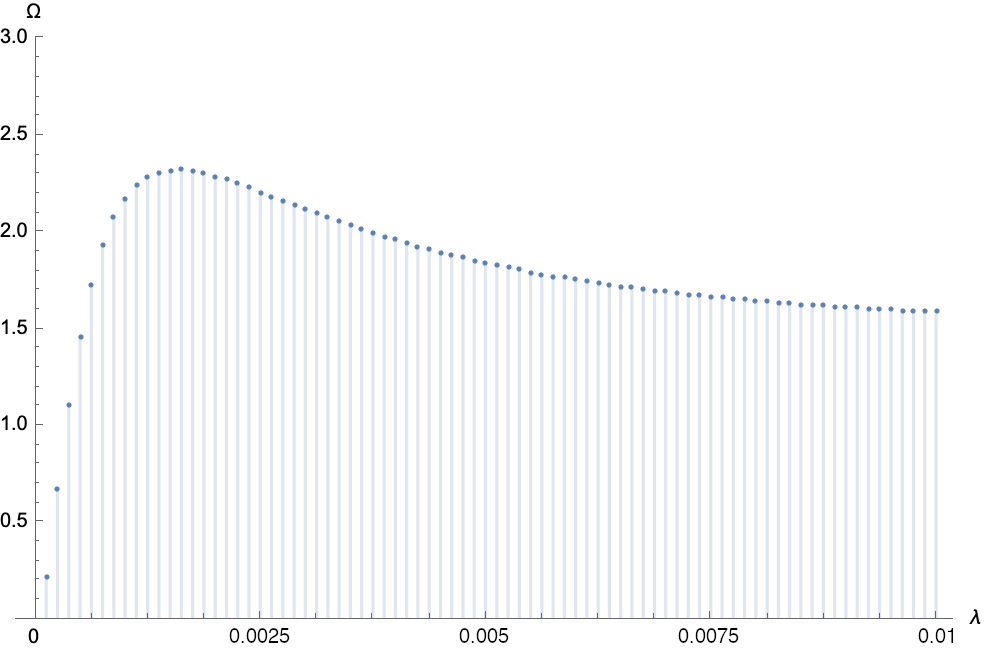}
    \caption{Plot of $\Omega$ against the flow parameter $\lambda$.}
    \label{sssoss}
\end{figure}
In figure \ref{sssoss}, it can be observed that $\Omega$ approaches the constant asymptotic value $\Omega_\infty\sim 1.52$.

\subsection{Scalar field with quartic potential}\label{quart}
\begin{figure}[H]
    \centering
    \includegraphics[width=0.8\linewidth]{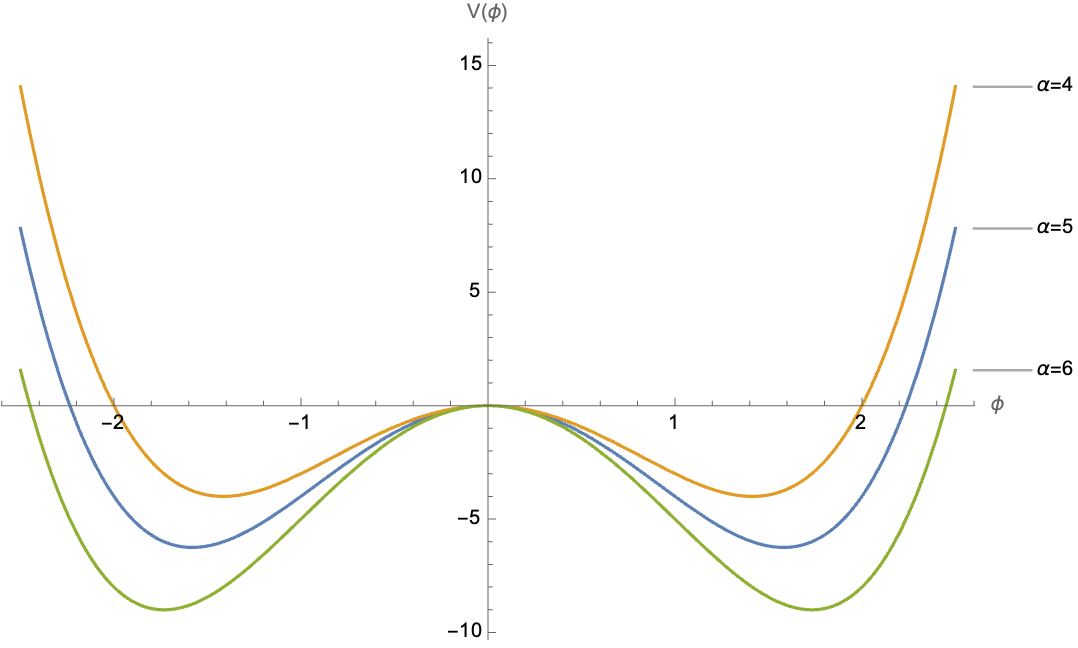}
    \caption{Plot of the behaviour of $V\left(\phi\right)$ around $\phi=0$, for three different values of $\alpha$.}
    \label{pottVg}
\end{figure}

\begin{figure}[H]
    \centering
    \includegraphics[width=0.8\linewidth]{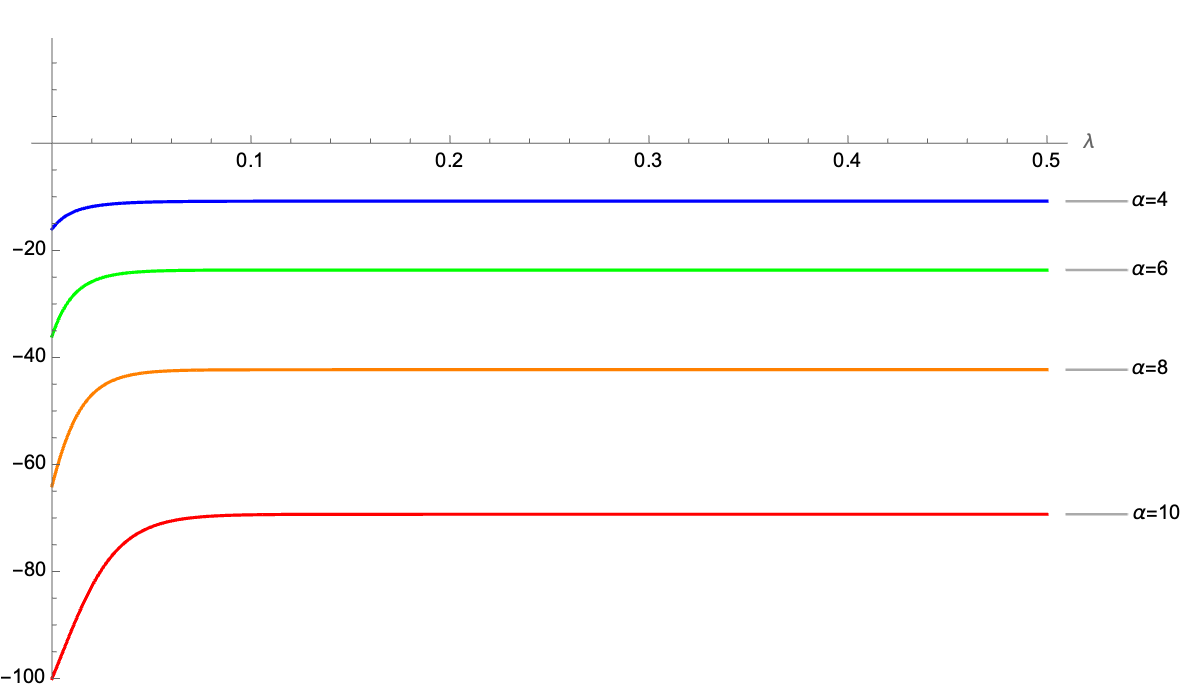}
    \caption{Plot of the flow behaviour of $R$ for different values of $\alpha$.}
    \label{mulR}
\end{figure}

\begin{figure}[H]
    \centering
    \includegraphics[width=0.8\linewidth]{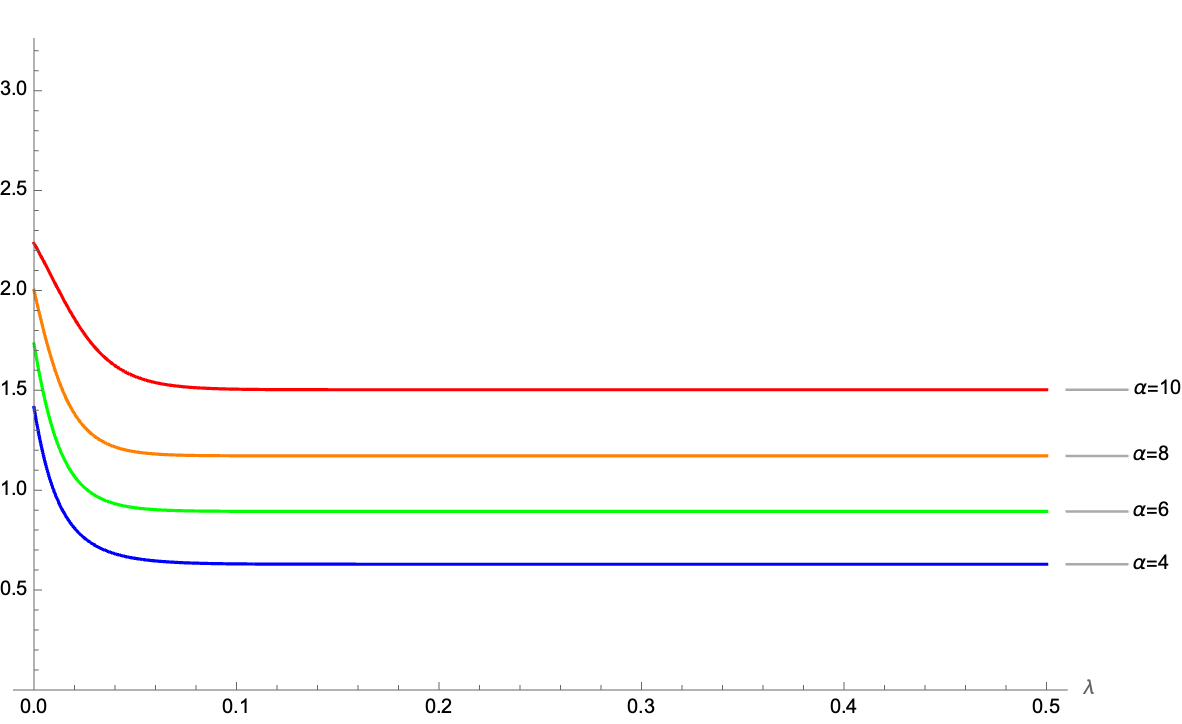}
    \caption{Plot of the flow behaviour of $\phi$ for different values of $\alpha$.}
    \label{mulp}
\end{figure}

\begin{figure}[H]
    \centering
    \includegraphics[width=0.9\linewidth]{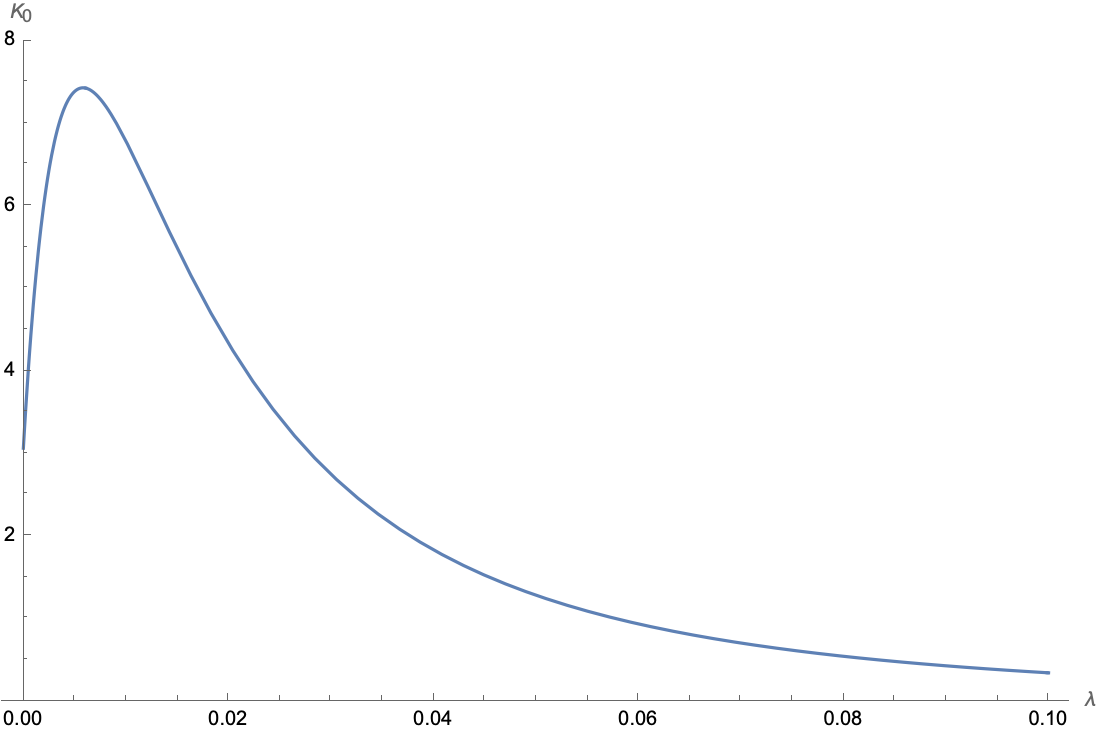}
    \caption{Plot of $K_0$ against the flow parameter $\lambda$, for $\alpha=1$.}
    \label{ssfsoss}
\end{figure}

In the following discussion, we assume to work with a single scalar field $\phi$, subject to a potential of the form
\be \label{6pot}
V\left(\phi\right)=\phi^4-\alpha\phi^2\ ,
\ee
with $\alpha$ being a positive constant to be fixed by the desired spacetime curvature initial value, which will indeed be a function of $\alpha$. The potential, for different values of $\alpha$, is shown in figure \ref{pottVg}.
At the minima corresponding to
\begin{equation}
    \phi_{\pm}=\pm\sqrt{\frac{\alpha}{2}}\ ,
\end{equation}
the potential takes the $\alpha$-dependent value:
\begin{equation}
    V\left(\phi_\pm\right)=-\frac{\alpha^2}{4}\ .
\end{equation}
Assuming the scalar field to take the constant $\phi_+$ or the constant value $\phi_-$ at the beginning of the flow, the potential produces an effective cosmological constant term
\be
\Lambda_{\text{eff}}=-\frac{\alpha^2}{4}\ .
\ee
Therefore, our initial metric will be that of $D$-dimensional Anti-de Sitter spacetime with scalar curvature:
\be 
R_0=-\frac{\alpha^2D}{2\left(D-2\right)}\ .
\ee
In $D=4$, which is the case which we are most interested in, we simply get:
\be 
R_0=-\alpha^2\ .
\ee
Therefore, our flow has the $\phi_+$ initial conditions
\begin{equation}
    R_0=-\alpha^2\ ,\quad \phi_+=\sqrt{\frac{\alpha}{2}}
\end{equation}
or the $\phi_-$ initial conditions:
\begin{equation}
    R_0=-\alpha^2\ ,\quad \phi_-=-\sqrt{\frac{\alpha}{2}}\ .
\end{equation}
From this point on, we choose the first one. Now, we derive the flow equations from the action as was done in \ref{fromact}. Neglecting the diffeomorphism term and observing that our initial conditions are spacetime-constant, the flow equations take the form
\begin{equation}\label{acteq}
    \begin{split}
        \frac{\de g_{\mu\nu}}{\de\lambda}=&-\left[\frac{R}{2}+G\left(\phi\right)\right]g_{\mu\nu}\ ,\\
        \frac{\de \phi}{\de\lambda}=&-2\left[\frac{R}{2}+G\left(\phi\right)\right]\ ,
    \end{split}
\end{equation}
where we have defined
\begin{equation}
    G\left(\phi\right)\equiv\sum_{n=1}^{+\infty}ns_{\ n}^{(4)}\phi^{n-1}
\end{equation}
via the introduction of the usual constants $s_{\ k}^{(4)}$:
\be 
    s_{\ k}^{(4)}\equiv\sum_{n=0}^{k}\frac{g_{n}}{\left(k-n\right)!n!}\left(-1\right)^{k-n}5^{n/2}\ .
\ee
For our specific potential, the only non-zero $g_n$ are:
\begin{equation}
    g_4=4!\ ,\quad g_2=-2\alpha\ .
\end{equation}
Therefore, we have:
\begin{equation}
    \begin{split}
       s_{\ 0}^{(4)}&=0\ ,\quad s_{\ 1}^{(4)}=0\ ,\quad s_{\ 2}^{(4)}=-5\alpha\ ,\\
       s_{\ 3}^{(4)}&=5\alpha\ ,\quad s_{\ k\geq 4}^{(4)}=-\frac{5\alpha}{\left(k-2\right)!}\left(-1\right)^{k-2}+\frac{25}{\left(k-4\right)!}\left(-1\right)^{k-4}\ .
    \end{split}
\end{equation}
Hence, we obtain:
\begin{equation}
\begin{split}
        G\left(\phi\right)&=-10\alpha\phi+15\alpha\phi^2-5\alpha\sum_{n=4}^{+\infty}\frac{n\phi^{n-1}\left(-1\right)^{n-2}}{\left(n-2\right)!}\\
        &\quad+25\sum_{n=4}^{+\infty}\frac{n\phi^{n-1}\left(-1\right)^{n-4}}{\left(n-4\right)!}\ .
\end{split}
\end{equation}
Considering the terms separately, we get:
\begin{equation}
\begin{split}
        \sum_{n=4}^{+\infty}\frac{n\phi^{n-1}\left(-1\right)^{n-2}}{\left(n-2\right)!}&=2\phi\sum_{n=2}^{+\infty}\frac{\phi^{n}\left(-1\right)^{n}}{n!}-\phi^2\sum_{n=1}^{+\infty}\frac{\phi^{n}\left(-1\right)^{n}}{n!}\\
        &=2\phi\left(e^{-\phi}+\phi-1\right)-\phi^2\left(e^{-\phi}-1\right)\\
        &=\phi\left(2-\phi\right) e^{-\phi}+\phi\left(3\phi-2\right)
\end{split}
\end{equation}
and, for the other one:
\begin{equation}
\begin{split}
        \sum_{n=4}^{+\infty}\frac{n\phi^{n-1}\left(-1\right)^{n-4}}{\left(n-4\right)!}&=\sum_{n=1}^{+\infty}\frac{n\phi^{n+3}\left(-1\right)^{n}}{n!}+4\sum_{n=0}^{+\infty}\frac{\phi^{n+3}\left(-1\right)^{n}}{n!}\\
        &=\phi^3\left(4-\phi\right)e^{-\phi}\ .
\end{split}
\end{equation}
Thus, we are left with:
\begin{equation}
\begin{split}
        G\left(\phi\right)=5\left[\left(20+\alpha\right)\phi^3-5\phi^4-2\alpha\phi^2\right]e^{-\phi}\ .
\end{split}
\end{equation}
We can, as in the previous section, write the flow equations as:
\begin{equation}
    \begin{split}
        \frac{\de R}{\de\lambda}&=-\frac{R}{2}\left[R+2G\left(\phi\right)\right]\ ,\\
        \frac{\de \phi}{\de\lambda}&=-\left[R+2G\left(\phi\right)\right]\ .
    \end{split}
\end{equation}
The flow behaviour for different values of $\alpha$ are captured in figures \ref{mulR} and \ref{mulp}.
Once more, we have
\begin{equation}
    S_{\mu\nu}=K_0g_{\mu\nu}\ ,
\end{equation}
where $K_0$ is defined as:
\begin{equation}
    K_0=-\phi\left(\phi^3+8\phi^2-\alpha\phi-4\alpha\right)\left[\frac{R}{2}+G\left(\phi\right)\right]\ .
\end{equation}
For any specific choice of $\alpha$, we can integrate $K_{0}$ in $\lambda$ to get the $\Gamma$ function in:
\begin{equation}
    \hat{T}_{\mu\nu}=\Gamma g_{\mu\nu}\ .
\end{equation}
In figure \ref{ssfsoss}, the explicit flow behaviour of $K_0$ is shown. There we can observe that it has positive values and then tends to zero. 
\begin{figure}[H]
    \centering
    \includegraphics[width=0.9\linewidth]{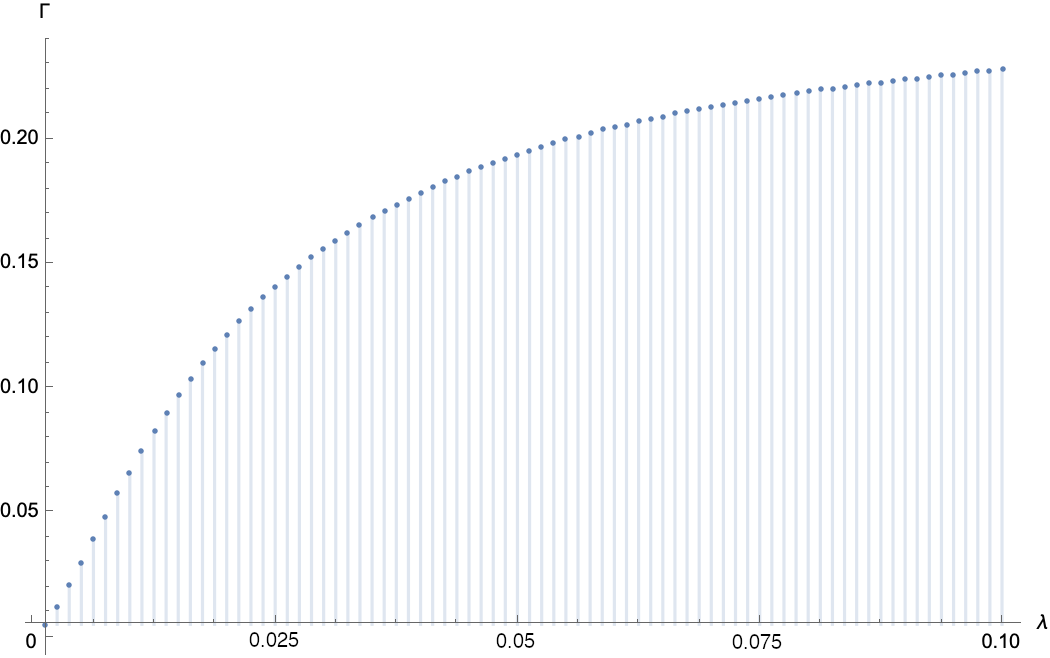}
    \caption{Plot of $\Gamma$ against the flow parameter $\lambda$, for $\alpha=1$.}
    \label{fsssoss}
\end{figure}
In figure \ref{fsssoss}, the explicit flow behaviour of $\Gamma$ is shown. As $\Omega$ did in \ref{fromact}, $\Gamma$ tends to a constant asymptotic value. For $\alpha=1$, it is $\Gamma_\infty\sim 0.24$. 
\subsection{Accounting for the extra term}
\begin{figure}[H]
    \centering
    \includegraphics[width=0.9\linewidth]{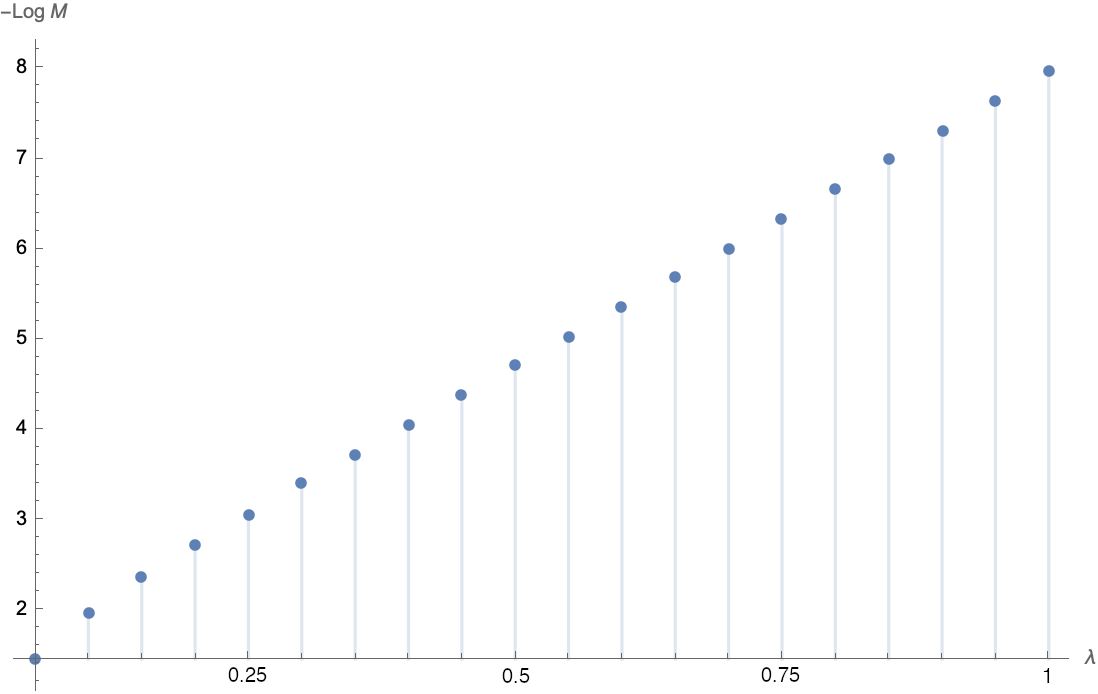}
    \caption{Plot of $-\log{M}$ against the flow parameter $\lambda$, for $\alpha=1$.}
    \label{logplot}
\end{figure}

In \ref{potentialcase} and \ref{quart}, when starting from a simple solution in $D=4$, flowing the metric and the scalar with the appropriate action-induced flow equations and imposing the system to be on-shell at any value of the flow parameter $\lambda$, we obtained that a new contribution $\hat{T}_{\mu\nu}$ to the energy-momentum tensor had to be introduced. In particular, by assuming it to be proportional to the metric as
\begin{equation}
    \hat{T}_{\mu\nu}=\frac{\hat{T}}{4}g_{\mu\nu}\ ,
\end{equation}
where $\hat{T}$ corresponds to $4\Omega$ in \ref{potentialcase} and to $4\Gamma$ in \ref{quart} respectively, we numerically derived its flow behaviour. Indeed, as can be clearly seen in figures \ref{sssoss} and \ref{fsssoss}, $\hat{T}$ approached an asymptotic, positive, finite value in both cases. If we want this feature to be accounted for by the appearance of some extra fields, the most natural way to achieve it is the following. Let's assume the extra fields to be labelled as $\varphi_k$ and to sit at the minimum of some potential $\hat{V}$, in which we include mass terms for the states, so that it takes the value
\be 
\hat{V}\left(\lambda\right)=\hat{V}_{\text{min}}+\sum_{k=1}^K\frac{m^2_k\left(\lambda\right)}{2}\ ,
\ee
where the whole flow dependence was moved to the masses. Now, let's assume all the $\varphi_k$ to be spacetime constant. They contribute to $\hat{T}_{\mu\nu}$ as:
\be 
\hat{T}_{\mu\nu}\left(\lambda\right)=\hat{V}\left(\lambda\right)g_{\mu\nu}\ .
\ee
Therefore, we must have:
\be \label{equivvvv}
\hat{V}\left(\lambda\right)=\frac{\hat{T}\left(\lambda\right)}{4}\ .
\ee
Since the fields do not appear in the theory at $\lambda=0$, their masses should be high at the initial point of the flow. This straightforwardly implies $\hat{V}_{\text{min}}$ to be negative, so that $\hat{V}\left(0\right)=0$ as expected from \eqref{equivvvv}. Thus, for $\hat{T}\left(\lambda\right)$ to have the flow behaviour derived in our discussions, we must take the masses contribution to $\hat{V}\left(\lambda\right)$ to decrease with $\lambda$. Namely, we have a tower of masses getting lighter with the flow. Concerning the specific value of $\hat{V}_{\text{min}}$, we can read it off from the asymptotic values $-\Omega_\infty\sim -1.52$ and $-\Gamma_\infty\sim -0.24$, as the masses contribution drops to zero. Therefore, the requirement of the equations of motion for the metric to be on-shell while following the action-induced geometric flow equations forces a tower of states to appear in the low energy theory. This is consistent with the naive prediction of a reasoning qualitatively tied to the Swampland Distance Conjecture. No statement can be made on the flow behaviour of the specific mass of one of such states, since they collectively realise $\hat{T}$. Nevertheless, an informed guess is within our reach. We consider the case of $\Gamma$ discussed in \ref{quart}, namely of Anti-de Sitter spacetime, and assume the masses to be factorised as
\be
m_k\left(\lambda\right)=f\left(k\right)\cdot M\left(\lambda\right)\ ,
\ee
so that the potential reduces to:
\be 
\hat{V}\left(\lambda\right)=\hat{V}_{\text{min}}+\gamma_0M^2\left(\lambda\right)\ ,
\ee
In the above expression, we have introduced the positive quantity:
\be
\gamma_0\equiv\sum_{k=1}^K\frac{f^2\left(k\right)}{2}\ .
\ee
Now, the flow behaviour of $M$ can be expressed as
\begin{equation}
    M\left(\lambda\right)=\sqrt{\frac{\hat{V}\left(\lambda\right)-\hat{V}_{\text{min}}}{\gamma_0}}\sim\sqrt{\hat{V}\left(\lambda\right)-\hat{V}_{\text{min}}}\ ,
\end{equation}
where the $1/\sqrt{\gamma_0}$ was neglected, since it is a numerical factor and we are interested in a qualitative analysis. By doing so, we can plot the flow behaviour of $-\log{M}$, as done in figure \ref{logplot}, and observe that it is approximately linear in $\lambda$. This precisely corresponds to the exponential drop of $M\left(\lambda\right)$ one would have expected from the Swampland Distance Conjecture.

\newpage
\section{Conclusions}
After a brief introduction, a general kind of $D$-dimensional gravity theory coupled to matter fields was presented in section \ref{theory}, in order to properly outline the boundaries inside which the whole discussion would have then developed. After having reviewed some remarkable and well-known properties of Ricci flow -and generalisations thereof- in section \ref{rfsec}, in which particular attention was dedicated to the problems that might arise when considering a manifold with Lorentzian signature or when moving to Ricci-Bourguignon flow, the issue of preserving the equations of motion for the metric tensor along a geometric flow was addressed. In section \ref{ossec}, general geometric flow equations were assumed for the metric and the cosmological constant. By enforcing Einstein field equations at any value of the flow parameter, an induced flow equation for the energy-momentum tensor was derived. It must be once more stressed that such a procedure neglected the details of the matter content of the theory, so that the energy-momentum tensor was consider as a generic, properly-behaved source for spacetime curvature. Therefore, the idea of finding a specific matter content, together with an appropriate set of geometric flow equations, that would have preserved both the metric and the matter fields on-shell along the flow was regarded as beyond the scope of the current work. It goes without saying that trying to solve such problem would be a great advancement with respect to the current state of our investigation. At that point, the main features of the energy-momentum tensor flow were studied, specifically for what concerns the case in which the metric evolves according to Ricci-Bourguignon flow. With the aim of connecting our analysis to the one developed in \cite{DeBiasio:2022omq}, $\mathcal{F}$-entropy functional were introduced in section \ref{entropysection} as the natural origin of geometric flow equations in the form of gradient flows. There, it was noted that, given a specific set of matter fields, one should expect the flow of the associated energy-momentum tensor induced by $\mathcal{F}$-entropy functional not to match the one obtained in section \ref{ossec}, for the corresponding evolution of the metric. Hence, it was suggested that the apparent discrepancy should be accounted for as the hint of new energy-momentum contributions appearing along the flow. This perspective was motivated by the Swampland distance conjecture, that claims large displacements in fields space must be accompanied by the emergence of tower of states in the low energy effective theory. This was extensively explored in section \ref{scaler}, where the case of a single spacetime scalar was considered. After having developed a general discussion, some simple examples were addressed explicitly either in the case of a free field or when the scalar is subject to a particular potential. Albeit still being an extremely simplified setting, the last one allowed us to highlight the main features one would expect if the Distance Conjecture was taken to hold true. Namely, we found that states whose masses exponentially drop along the flow represent the most natural candidate for the extra energy-momentum tensor terms appearing due to action-induced geometric flow equations. 
\newpage

\appendix\section{Energy-momentum flow with scalars}\label{compute}
In the following appendix, we explicitly compute 
\be
\begin{split}
   2B_{\mu\nu}&=\nabla^\sigma\nabla_\nu A_{\mu\sigma}+\nabla_\mu\nabla^\sigma A_{\sigma\nu} -\nabla_\mu\nabla_\nu A-R_\mu{}^\sigma{}_\nu {}^\theta A_{\sigma\theta}\\
       &\quad+2C g_{\mu\nu}-\Delta A_{\mu\nu }+R_\mu{}^\sigma A_{\sigma\nu }-A_{\mu\nu}R+2\Lambda A_{\mu\nu}\\
  &\quad 
  -g_{\mu\nu}\nabla^\sigma\nabla^\theta A_{\sigma\theta}+g_{\mu\nu}\Delta A+g_{\mu\nu}A^{\sigma\theta}R_{\sigma\theta}
\end{split}
\ee
when the expression for $A_{\mu\nu}$ is the one postulated in \eqref{mss}:
\be
\begin{split}
     A_{\mu\nu}&\equiv -2R_{\mu\nu}+2\vartheta Rg_{\mu\nu}+\sum_{j=1}^N b_j\nabla_\mu\nabla_\nu\varphi_j+g_{\mu\nu}\sum_{j=1}^N d_j\Delta\varphi_j\\
     &\quad+\sum_{i,j=1}^N e_{ij}\nabla_\mu\varphi_i\nabla_\nu\varphi_j+g_{\mu\nu}\sum_{i,j=1}^N f_{ij}\nabla_\alpha\varphi_i\nabla^\alpha\varphi_j\ .
\end{split}
\ee 
Its trace, therefore, takes the form:
\begin{equation}
    A=2\left(D\vartheta-1\right)R+\sum_{j=1}^N\left(Dd_j+b_j\right)\Delta\varphi_j+\sum_{i,j=1}^N\left(Df_{ij}+e_{ij}\right)\nabla_\alpha\varphi_i\nabla^\alpha\varphi_j\ .
\end{equation}
In order to do so, we split the expression in four different terms:
\be 
\begin{split}
     M_{\mu\nu}&=\nabla^\sigma\nabla_\nu A_{\mu\sigma}+\nabla_\mu\nabla^\sigma A_{\sigma\nu} -\nabla_\mu\nabla_\nu A-\Delta A_{\mu\nu }\ ,\\
     N_{\mu\nu}&=R_\mu{}^\sigma A_{\sigma\nu }-R_\mu{}^\sigma{}_\nu {}^\theta A_{\sigma\theta}-A_{\mu\nu}R\ ,\\
     O_{\mu\nu}&=g_{\mu\nu}\Delta A-g_{\mu\nu}\nabla^\sigma\nabla^\theta A_{\sigma\theta}\ ,\\
     P_{\mu\nu}&=2C g_{\mu\nu}+2\Lambda A_{\mu\nu}+g_{\mu\nu}A^{\sigma\theta}R_{\sigma\theta}\ .
\end{split}
\ee
Now, we start analysing them one by one. For $M_{\mu\nu}$, we get:
\be
\begin{split}
    M_{\mu\nu}&=-2\nabla^\sigma\nabla_\nu R_{\mu\sigma}+2\left[1+\left(2-D\right)\vartheta\right]\nabla_\mu\nabla_\nu R+2\Delta R_{\mu\nu}\\
&\quad-2\vartheta g_{\mu\nu} \Delta R-2\nabla_\mu\nabla^\sigma R_{\sigma\nu}+\sum_{j=1}^N b_j\nabla_\mu\nabla^2\nabla_\nu\varphi_j\\
&\quad+\sum_{j=1}^N\left[\left(2-D\right)d_j-b_j\right]\nabla_\mu\nabla_\nu\Delta\varphi_j-\sum_{j=1}^N b_j\Delta\nabla_\mu\nabla_\nu\varphi_j\\
&\quad+\sum_{j=1}^N b_j\nabla^\sigma\nabla_\nu \nabla_\mu\nabla_\sigma\varphi_j-g_{\mu\nu}\sum_{j=1}^N d_j\Delta\Delta\varphi_j\\
&\quad+\nabla^\sigma\nabla_\nu\sum_{i,j=1}^N e_{ij}\nabla_\mu\varphi_i\nabla_\sigma\varphi_j+\nabla_\mu\nabla_\nu\sum_{i,j=1}^N f_{ij}\nabla_\alpha\varphi_i\nabla^\alpha\varphi_j\\
    &\quad+\nabla_\mu\nabla^\sigma \sum_{i,j=1}^N e_{ij}\nabla_\sigma\varphi_i\nabla_\nu\varphi_j+\nabla_\mu\nabla_\nu \sum_{i,j=1}^N f_{ij}\nabla_\alpha\varphi_i\nabla^\alpha\varphi_j\\
    &\quad-\Delta\sum_{i,j=1}^N e_{ij}\nabla_\mu\varphi_i\nabla_\nu\varphi_j-g_{\mu\nu}\Delta\sum_{i,j=1}^N f_{ij}\nabla_\alpha\varphi_i\nabla^\alpha\varphi_j\\
        &\quad-\nabla_\mu\nabla_\nu\sum_{i,j=1}^N\left(Df_{ij}+e_{ij}\right)\nabla_\alpha\varphi_i\nabla^\alpha\varphi_j\ .
\end{split}
\ee 
Concerning, on the other hand, $N_{\mu\nu}$, we obtain:
\be
\begin{split}
    N_{\mu\nu}&=\sum_{j=1}^N\left[\left(R_\mu{}^\sigma\delta_\nu{}^{\theta}-R_\mu{}^\sigma{}_\nu {}^\theta-R\delta_\mu{}^\sigma\delta_\nu{}^{\theta}\right) b_j\nabla_\sigma\nabla_\theta\varphi_j-Rg_{\mu\nu}d_j\Delta\varphi_j\right]\\
    &\quad+2R_\mu{}^\sigma{}_\nu {}^\theta R_{\sigma\theta}+2RR_{\mu\nu}-2\vartheta R^2g_{\mu\nu}-2R_\mu{}^\sigma R_{\sigma\nu}\\
    &\quad+R_\mu{}^\sigma\sum_{i,j=1}^N e_{ij}\nabla_\sigma\varphi_i\nabla_\nu\varphi_j+R_{\mu\nu}\sum_{i,j=1}^N f_{ij}\nabla_\alpha\varphi_i\nabla^\alpha\varphi_j\\
&\quad-R_\mu{}^\sigma{}_\nu {}^\theta\sum_{i,j=1}^N e_{ij}\nabla_\sigma\varphi_i\nabla_\theta\varphi_j-R_{\mu\nu}\sum_{i,j=1}^N f_{ij}\nabla_\alpha\varphi_i\nabla^\alpha\varphi_j\\
&\quad-R\sum_{i,j=1}^N e_{ij}\nabla_\mu\varphi_i\nabla_\nu\varphi_j-g_{\mu\nu}R\sum_{i,j=1}^N f_{ij}\nabla_\alpha\varphi_i\nabla^\alpha\varphi_j\ .
\end{split}
\ee 
For $P_{\mu\nu}$, the result is:
\be
\begin{split}
    P_{\mu\nu}&=\sum_{j=1}^N \left[b_j\left(2\Lambda\delta_\mu{}^\sigma\delta_\nu{}^{\theta}+g_{\mu\nu}R^{\sigma\theta}\right)\nabla_\sigma\nabla_\theta\varphi_j+d_jg_{\mu\nu}\left(2\Lambda+R\right)\Delta\varphi_j\right]\\
    &\quad+2\left(C+\vartheta R^2-R_{\sigma\theta}R^{\sigma\theta}+2\Lambda\vartheta R\right)g_{\mu\nu}-4\Lambda R_{\mu\nu}\\
    &\quad+2\Lambda \sum_{i,j=1}^N e_{ij}\nabla_\mu\varphi_i\nabla_\nu\varphi_j+2\Lambda g_{\mu\nu} \sum_{i,j=1}^N f_{ij}\nabla_\alpha\varphi_i\nabla^\alpha\varphi_j\\
 &\quad+g_{\mu\nu}\sum_{i,j=1}^N e_{ij}R_{\sigma\theta}\nabla^\sigma\varphi_i\nabla^\theta\varphi_j+g_{\mu\nu}R\sum_{i,j=1}^N f_{ij}\nabla_\alpha\varphi_i\nabla^\alpha\varphi_j\ .
\end{split}
\ee 
For the term we labelled $O_{\mu\nu}$, we get the expression:
\be
\begin{split}
    O_{\mu\nu}&=g_{\mu\nu}\sum_{j=1}^N\left\{\left[\left(D-1\right)d_j+b_j\right]\Delta\Delta\varphi_j-b_j\nabla^\sigma\nabla^\theta\nabla_\sigma\nabla_\theta\varphi_j\right\}\\
    &\quad+\left[2\left(D-1\right)\vartheta-1\right]g_{\mu\nu}\Delta R+g_{\mu\nu}\Delta\sum_{i,j=1}^N\left(Df_{ij}+e_{ij}\right)  \nabla_\alpha\varphi_i\nabla^\alpha\varphi_j\\
    &\quad-g_{\mu\nu}\nabla^\sigma\nabla^\theta \sum_{i,j=1}^N e_{ij}\nabla_\sigma\varphi_i\nabla_\theta\varphi_j-g_{\mu\nu}\Delta\sum_{i,j=1}^N f_{ij}\nabla_\alpha\varphi_i\nabla^\alpha\varphi_j\ .
\end{split}
\ee 
Combining all the factors, we are left with:
\be 
\begin{split}\label{souceone}
    2B_{\mu\nu}&=-2\nabla^\sigma\nabla_\nu R_{\mu\sigma}+2\left[1+\left(2-D\right)\vartheta\right]\nabla_\mu\nabla_\nu R+2\Delta R_{\mu\nu}\\
&\quad-2\nabla_\mu\nabla^\sigma R_{\sigma\nu}+\left[2\left(D-2\right)\vartheta-1 \right]g_{\mu\nu}\Delta R\\
    &\quad+2R_\mu{}^\sigma{}_\nu {}^\theta R_{\sigma\theta}+2RR_{\mu\nu}-2R_\mu{}^\sigma R_{\sigma\nu}-4\Lambda R_{\mu\nu}\\
    &\quad+2\left(C-R_{\sigma\theta}R^{\sigma\theta}+2\Lambda\vartheta R\right)g_{\mu\nu}+\sum_{j=1}^N\Omega^{\ j}_{\mu\nu}\varphi_j\\
        &\quad     +\sum_{i,j=1}^N\Theta_{\mu\nu}^{\ ij}\nabla_\alpha\varphi_i\nabla^\alpha\varphi_j+\sum_{i,j=1}^N\Sigma_{\mu\nu}^{\sigma\theta ij}\nabla_\sigma\varphi_i\nabla_\theta\varphi_j\ .
\end{split}
\ee
Where we have introduced $N$ metric-dependent differential operators
\be 
\begin{split}
    \Omega^{\ j}_{\mu\nu}&= b_j\nabla_\mu\nabla^2\nabla_\nu-\left[\left(D-2\right)d_j+b_j\right]\nabla_\mu\nabla_\nu\Delta- b_j\Delta\nabla_\mu\nabla_\nu\\
    &\quad+ \left[b_j\left(2\Lambda\delta_\mu{}^\sigma\delta_\nu{}^{\theta}+g_{\mu\nu}R^{\sigma\theta}\right)\nabla_\sigma\nabla_\theta\right]+ b_j\nabla^\sigma\nabla_\nu \nabla_\mu\nabla_\sigma\\
    &\quad+g_{\mu\nu}\left\{\left[\left(D-2\right)d_j+b_j\right]\Delta\Delta-b_j\nabla^\sigma\nabla^\theta\nabla_\sigma\nabla_\theta+2d_j\Lambda\Delta\right\}\\
    &\quad+\left[\left(R_\mu{}^\sigma\delta_\nu{}^{\theta}-R_\mu{}^\sigma{}_\nu {}^\theta-R\delta_\mu{}^\sigma\delta_\nu{}^{\theta}\right) b_j\nabla_\sigma\nabla_\theta\right]
\end{split}
\ee
and $2N^2$ metric-dependent differential operators:
\be
\Theta_{\mu\nu}^{\ ij}\equiv\left(g_{\mu\nu}\Delta-\nabla_\mu\nabla_\nu \right)\left(Df_{ij}+e_{ij}\right)+2\left(\nabla_\mu\nabla_\nu-g_{\mu\nu}\Delta+\Lambda g_{\mu\nu}\right)f_{ij}\ ,
\ee
\be
\begin{split}
   \Sigma_{\mu\nu}^{\sigma\theta ij}&\equiv \Big[\left(R_\mu{}^\sigma+\nabla_\mu\nabla^\sigma \right)\delta_\nu^{\ \theta}+g_{\mu\nu}R^{\sigma\theta}-g_{\mu\nu}\nabla^\sigma\nabla^\theta -R_\mu{}^\sigma{}_\nu {}^\theta\\
&\quad+\left(2\Lambda-R-\Delta \right)\delta_{\mu}^{\ \sigma}\delta_{\nu}^{\ \theta}+\delta_{\mu}^{\ \sigma}\nabla^\theta\nabla_\nu\Big]e_{ij} \ .
\end{split}
\ee
If we impose $d_j=b_j=e_{ij}=f_{ij}=0$, \eqref{souceone} consistently matches with \eqref{rbosB}.

\section{Covariant derivatives in $D=2$}\label{app2d}
Given the setting outlined in \ref{specific}, with the metric
\be 
g_{\mu\nu}=\begin{pmatrix}
-g_1\left(x\right) & 0 \\
a & g_2\left(x\right)
\end{pmatrix}
\ee
and the Christoffel symbols:
\be 
\Gamma^t_{\ tx}=\Gamma^t_{\ xt}=\frac{g_1'}{2g_1}\ ,\quad\Gamma^x_{\ tt}=\frac{g_1'}{2g_2}\ ,\quad\Gamma^x_{\ xx}=\frac{g_2'}{2g_2}\ .
\ee 
Taking a time-independent scalar $\phi$, we compute its non-zero covariant derivatives as:
\be
\nabla_x\phi=\phi'\ .
\ee
\be
\begin{split}
    \nabla_t\nabla_t\phi&=-\frac{g_1'}{2g_2}\phi'\ ,\quad \nabla_x\nabla_x\phi=\phi''-\frac{g_2'}{2g_2}\phi'
\end{split}
\ee

\be
\begin{split}
    \nabla_t\nabla_t\nabla_x\phi &=-\frac{g_1'}{2g_2}\phi''+\left[\frac{g_1'g_2'}{4g_2^2}+\frac{\left(g_1'\right)^2}{4g_1g_2}\right]\phi'
\end{split}
\ee

\be
\begin{split}
    \nabla_t\nabla_x\nabla_t\phi &=-\frac{g_1'}{2g_2}\phi''+\left[\frac{g_1'g_2'}{4g_2^2}+\frac{\left(g_1'\right)^2}{4g_1g_2}\right]\phi'
\end{split}
\ee

\be
\begin{split}
    \nabla_x\nabla_t\nabla_t\phi &=-\frac{g_1'}{2g_2}\phi''+\left[\frac{\left(g_1'\right)^2}{2g_1g_2}-\frac{2g_1''g_2-2g_1'g_2'}{4g_2^2}\right]\phi'
\end{split}
\ee

\be
\begin{split}
    \nabla_x\nabla_x\nabla_x\phi &=\phi'''-\frac{3g_2'}{2g_2}\phi''+\left[2\frac{\left(g_2'\right)^2}{4g_2^2}-\frac{2g_2''g_2-2\left(g_2'\right)^2}{4g_2^2}\right]\phi'
\end{split}
\ee

\newpage
\bibliographystyle{utphys}
\bibliography{bibliogra.bib}
\end{document}